\documentclass[11pt,a4paper]{article}
\usepackage{amsmath}
\usepackage{enumerate}
\usepackage{amssymb}
\usepackage{graphicx}
\usepackage{multirow}
\usepackage{xcolor}
\usepackage{tablefootnote}
\usepackage{threeparttable}
\usepackage{bm}
\usepackage[normalem]{ulem}
\usepackage{jcappub}

\begin{document}

\title{Inferring the Age of the Universe with Globular Clusters}
\author[a,b]{David Valcin,}
\author[c,a,b]{Jos\'e Luis Bernal,}
\author[a,d]{Raul Jimenez,}
\author[a,d]{Licia Verde,}
\author[e,f,g]{Benjamin D.~Wandelt}

\affiliation[a]{ICC, University of Barcelona, IEEC-UB, Mart\' i i Franqu\` es, 1, E08028
Barcelona, Spain}
\affiliation[b]{Dept. de F\' isica Qu\` antica i Astrof\' isica, Universitat de Barcelona, Mart\' i i Franqu\` es 1, E08028 Barcelona,
Spain}
\affiliation[c]{Department of Physics and Astronomy, Johns Hopkins University, 3400 North Charles Street, Baltimore, Maryland 21218, USA}
\affiliation[d]{ICREA, Pg. Lluis Companys 23, Barcelona, 08010, Spain.} 
\affiliation[e]{Sorbonne Universit\'e, CNRS, UMR 7095, Institut d'Astrophysique de Paris, 98 bis bd Arago, 75014 Paris, France.}
\affiliation[f]{Sorbonne Universit\'e, Institut Lagrange de Paris (ILP), 98 bis bd Arago, 75014 Paris, France.}
\affiliation[g]{Center for Computational Astrophysics, Flatiron Institute, 162 5th Avenue, 10010, New York, NY, USA.}

\emailAdd{d.valcin@icc.ub.edu}
\emailAdd{jbernal2@jhu.edu} 
\emailAdd{raul.jimenez@icc.ub.edu}
\emailAdd{liciaverde@icc.ub.edu}
\emailAdd{bwandelt@iap.fr}

\abstract{We present an estimate of the absolute age of 68  galactic globular clusters obtained by  exploiting the distribution of stars in the  full color-magnitude diagram.  In particular, we jointly estimate the absolute age, distance, reddening, metallicity ([Fe/H]) and [$\alpha$/Fe] of each cluster, imposing  priors motivated by independent observations; we also estimate possible systematics from stellar modeling. Our derived distances for the globular cluster sample are in agreement with those obtained from GAIA using main-sequence dwarf stars (where available), and the inferred ages are in  good agreement with those previously published. The novelty of our approach is that, with the adopted priors, we are able to estimate robustly these parameters  from the globular cluster color-magnitude diagram. We find that the average age of the oldest globular clusters is  $t_{\rm GC}=13.32 \pm 0.1 {\rm (stat.)} \pm 0.5 {\rm (sys.)}$, at 68\% confidence level, including systematic uncertainties from stellar modeling. These measurements can be used to infer the age of the Universe, largely independently of the cosmological parameters: we find an age of the Universe  $t_{\rm U}=13.5^{+0.16}_{-0.14} {\rm (stat.)} \pm 0.5 ({\rm sys.})$ at 68\% confidence level, accounting for the formation time of globular clusters and its uncertainty. This value is compatible with $13.8 \pm 0.02$ Gyr, the cosmological model-dependent value inferred by the Planck mission assuming the $\Lambda$CDM model.}

\maketitle

\section{Introduction}
\label{sec:intro}

The color  magnitude diagram of co-eval stellar populations in the Milky Way can be used to infer  the age of its oldest stars. The age can also be estimated  for individual stars if their metallicity and the distance to them are  known. For resolved stellar populations,  however, an independent measurement of the distance is not strictly necessary as  
the full morphology of the color-magnitude diagram can, in principle, provide a determination of the absolute age. There is extensive literature on this subject; reviews can be found in e.g.,  Refs.~\cite{Catelan,Soderblom,Bolte+}.

Historically, the age of the oldest stellar populations in the Milky Way has been measured using the luminosity of the main-sequence turn off  point (MSTOP) in the color-magnitude diagram of globular clusters (GCs). Globular clusters are (almost-- more on this below) single stellar populations of stars (see e.g., Ref.~\cite{Bolte+}). It has long been recognized that they are among the most metal poor  ($\sim 1 \%$ of the solar metallicity) stellar systems in the Milky Way, and exhibit color-magnitude diagrams characteristic of old ($> 10$ Gyr) stellar populations~\cite{OMalley,Catelan,Bolte+}. 

In fact, the first quantitative attempt to compute the age of the globular cluster M3 was made by Haselgrove and Hoyle more than 60 years ago~\cite{Hoyle}. In this work, stellar models were computed on the early Cambridge mainframe computer and its results compared ``by eye" to the observed color-magnitude diagram. A few stellar phases were computed by solving the equations of stellar structure; this output was compared to observations. Their estimated age for M3 is only 50\% off from its current value.\footnote{Their low age estimate is due to the use of an incorrect distance to M3, since the stellar model used deviated  just $\sim$ 10\% from current models' prediction of  the effective temperature and gravity of stars, with their same, correct assumptions~\cite{Bolte+}.} This was the first true attempt to use computer models to fit resolved stellar populations and thus obtain cosmological parameters: the age of the Universe in this case. Previous estimates of the ages of GCs involved just analytic calculations, which significantly impacted the accuracy of the results, given the complexity of the stellar structure equations (see e.g., Ref.~\cite{Sandage}).

The absolute age of a GC inferred using only the MSTOP luminosity is degenerate with other properties of the GC.  
As already shown in the pioneering work of  Ref.~\cite{Hoyle}, the distance uncertainty to the GC entails the largest contribution to the error budget: a given \% level of relative uncertainty in the distance determination involves roughly the same level of uncertainty in the inference of the age. Other sources of uncertainty are:  the metallicity content, the Helium fraction, the dust absorption  \cite{Bolte+} and theoretical systematics regarding the physics and modeling of stellar evolution. 

However, there is more information enclosed in the full-color magnitude diagram of a GC than that enclosed in its MSTOP.
As first pointed out in Refs.~\cite{JimenezPadoanLF,PadoanJimenezLF}, the full color-magnitude diagram has features that allow for a joint fit of the distance scale and the age (see Appendix \ref{sec:sensitivity} for a visual rendering of this). On the one hand, figure~2 in Ref.~\cite{JimenezPadoanGC} shows how the different portions of the color-magnitude diagram constrain the corresponding physical quantities. On the other, figure~1 in Ref.~\cite{PadoanJimenezLF} and figure~3 in Ref.~\cite{JimenezPadoanGC} show how the luminosity function is not a pure power law but has features that contain information about the different physical parameters of the GC. This technique enabled the estimation of the ages of the GCs M68~\cite{JimenezPadoanLF}, M5 and M55~\cite{JimenezPadoanGC}. Moreover, in principle,  exploiting the morphology of the horizontal branch makes it possible to determine the ages of GCs independently of the distance~\cite{JimenezGC96}.

Further, on the observational front, the gathering of Hubble Space Telescope (HST)  photometry for a significant sample of galactic GCs has been a game changer. HST has provided  very accurate photometry with a very compact point spread function, thus easing the problems of crowding when attempting to extract the color-magnitude diagram for a GC and  making it  much easier to control contamination from  foreground and background field stars. 

For these reasons, a precise and robust determination of the age of a GC requires a global fit of all these quantities from the full color-magnitude diagram  of the cluster. In order to exploit this information, and due to degeneracies among GC parameters, we need a suitable statistical approach.  Bayesian techniques, which  have recently become the workhorse of  cosmological parameter inference, are of particular interest.  
 In the perspective of  possibly using  the estimated  age of the oldest stellar populations  in a cosmological context as a route to constrain  the age of the Universe, it is of  value to adopt Bayesian techniques in this context too.
  
There are only a few recent attempts at using Bayesian techniques  to fit GCs' color-magnitude diagrams, albeit only using some of their features (see e.g., Ref.~\cite{BayesianGC}). Other attempts to use Bayesian techniques to age-date individual stars from the GAIA catalog can be found in Ref.~\cite{Lund}. A limitation with the methodology presented in Ref~\cite{BayesianGC} is the large number of parameters  needed in their likelihood. Actually, for a GC of $N _{\rm stars}$ there are, in principle, $4 \times N_{\rm stars} + 5$ model's parameters (effectively $3 \times N_{\rm stars} + 5$), where the variables for each star are: initial stellar mass, photometry, ratio of secondary to primary initial stellar masses (fixed to 0 in Ref.~\cite{BayesianGC}) and  cluster membership indicator. In addition,  there are 5 (4) additional GC variables, namely: age, metallicity (fixed in the analysis of Ref. ~\cite{BayesianGC}), distance modulus, absorption and Helium fraction. For a cluster of 10,000 or more stars, the computational cost of this approach is very high. To overcome this issue Ref.~\cite{BayesianGC} randomly selected a subsample of 3000 stars, half above  and half below the MSTOP of the cluster,  ``to ensure a reasonable sample of stars on the sub-giant and red-giant branches". Another difficulty arises from the fact that  the cluster membership indicator  variable  can take only the value of 0 or 1 (i.e., whether a star belongs to the cluster or not). This creates a sample of two populations referred  to as a  \emph{finite mixture distributions}~\cite{BayesianGC}. 

Capitalizing on  the wide availability and potential of current observations, the aim of this paper is to present  a  Bayesian approach  to exploit features in the color-magnitude diagram  beyond the  MSTOP and determine robustly the absolute age, jointly with all other relevant quantities such as metallicity, distance, dust absorption and abundance of $\alpha$-enhanced elements, of each GC. In addition to statistical errors, we estimate systematic theoretical uncertainties regarding the stellar model. We bypass the computational challenge of the approach explored in Ref. \cite{BayesianGC} by introducing some simplifications and by coarse-graining the information in the  GC color-magnitude diagram, which greatly reduces the dimensionality of the problem without significant loss of information.

Our paper is organized as follows. In \S~\ref{sec:data} we describe the HST GC data;  the stellar model used to fit the data and the calibration of the GC data is shown in \S\ref{sec:calib}. The approach  developed to obtain the parameters of GCs is introduced in \S~\ref{sec:inference} where we describe the likelihood adopted and how we explore the posterior with Monte Carlo Markov chains. Results, the age of the oldest GCs and the corresponding inferred age of the Universe are presented in \S~\ref{sec:results}. We expose our conclusions in \S~\ref{sec:summary}. A series of appendixes cover the  technical details of our method.

\section{Data and stellar model}
\label{sec:data}

\subsection{Globular cluster catalogs: defining our sample}

We use the HST-ACS catalog of 65 globular clusters~\cite{Sarajedini2007} plus $6$ additional ones from Ref.~\cite{Dotter2010}. Out of 71 clusters, two were removed because of high differential reddening and a lack of red giant branch stars~\cite{BayesianGC}, one more was removed because of a lack of reasonable extinction prior from the literature,  leaving 68 clusters in total. The data are available in two different Vega filters: F606W and F814W.

In order to clean the data of stars with poorly determined photometry, we use the same  prescriptions as in Ref.~\cite{BayesianGC}. First, we remove stars for which photometric errors,\footnote{Each photometric error has been rescaled  depending on the number of observations according to the catalog  instructions  in the {\tt readme} file.} in both filters, fall into the outer 5\% tail of the distribution. Then, we also remove stars in the outer 2.5\% tails of the distributions of X and Y pixel location errors.  Indeed,  large pixel location errors indicate a non-reliable measurement of the properties of the star.

Similarly, we also expect measurements to be less robust at very low magnitudes. Moreover, the photometric error corresponding to  these stars becomes very large, reducing drastically the information content of this part of the color-magnitude diagram.  

 Hence, for each cluster we define a  ``functional'' magnitude interval between the lowest apparent magnitude of the brightest stars and a magnitude cut arbitrarily defined at $m_{F606W} = 26$, to include most of the main sequence stars for every cluster.

 Only stars that satisfy all the conditions listed above and belong to the defined  functional magnitude interval are considered further.\footnote{A further cut at low magnitudes is introduced in Sec. ~3. The cut described here is motivated by the survey limitations; the cut in  Sec.~3 is  to speed up the analysis without removing significant signal.}   For readers interested in the number and percentage of stars retained, details are reported in Tab.~\ref{tab:stars_number} of appendix~\ref{app:GCtable}.

\subsection{Software and stellar models}
For the theoretical modeling of the data, we choose to work with a modified version of the software package \texttt{isochrones}\footnote{\url{https://github.com/timothydmorton/isochrones}, version 1.1-dev.}~\citep{isochrones}. This software reads synthetic photometry files provided by stellar models and then interpolates magnitudes along isochrones (points in the stellar evolutionary track at same age) correcting for absorption, given the input parameters. Even though a new version is currently under development (\texttt{isochrones}2.0),  and that in the main text  of this paper we only use one model, we decided to use a modified version of the previous release as it enables us  to consider  different stellar models.
The two stellar models already implemented are \texttt{MIST}~\cite{MIST0,MIST1} and  \texttt{DSED}~\cite{dsed}. Each  stellar model  comprises a set of {\it photometry files} that correspond to (discretized) isochrones in a color magnitude diagram. However,  it is important to note that only in the photometry files of \texttt{DSED} several different abundances (parameterised by  [$\alpha$/Fe]) of $\alpha$-enhanced elements, other than the solar abundance, are provided. These are elements like O, Ne, Mg, Si, S, Ca and Ti that are created via $\alpha-$particle  (helium nucleus) capture; 
 [$\alpha$/Fe] is fixed to  0 in the photometry files corresponding to the \texttt{MIST} model. This is important as GCs do have non-solar-scaled abundances. As we will show below (see  Appendix \ref{sec:sensitivity}) the abundance [$\alpha$/Fe] is partially (but only partially) degenerate with variations of the GC's age and metallicity, so that it must be considered as a free parameter in the analysis to avoid biasing the results and to infer the correct statistical uncertainties. Therefore, we  consider [$\alpha$/Fe] as an independent parameter and limit our analysis to the \texttt{DSED} model; the ranges in parameter space covered by the \texttt{DSED} model photometry files in \texttt{isochrones} are specified in  Tab.~\ref{tab:Table1}.

 \begin{table}[]
\centering
\begin{threeparttable}
\begin{tabular}{|c|c|}
\hline
Stellar model & DSED \\ \hline\hline
initial rotation rate $v/v_{crit}$ &  0.0 \\ \hline
Age range & 0.250-15 Gyr \\ \hline
Age sampling &  0.5 Gyr \\ \hline
number of EEPs per isochrone & $\simeq$ 270 \\ \hline
Metallicity range {[}Fe/H{]} & -2.5 to 0.5 dex\\ \hline
\begin{tabular}[c]{@{}c@{}}Helium fraction configuration\\ \end{tabular} &  $Y_{\rm init}$ = 0.245\tnote{\dag}, 0.33, 0.40 \tnote{\ddag} \\ \hline
 {[}$\alpha$/Fe{]} & -0.2 to 0.8\tnote{+} \\ \hline
\end{tabular}%
\footnotesize
\begin{tablenotes}
\item[\dag] The varying Helium fraction configurations, $Y$, are defined in photometry files as $Y = Y_{\rm init} + 1.5 Z$ where $Z$ is the metal mass fraction and $Y_{\rm init}$ is the starting value.
\item[\ddag] Fixed Helium fraction configurations $Y=0.33 $ and 0.40 are only available for [Fe/H] $\leq$ 0.
\item[+] For the fixed Helium fraction configurations, only two options [$\alpha$/Fe]$=0$ or $+0.4$ are available.
\end{tablenotes}
\end{threeparttable}
\caption{Properties of the  \texttt{DSED} stellar models available in the \texttt{isochrones} package. We refer the reader to original  Ref.~\cite{dsed} for more details.
}
\label{tab:Table1}
\end{table}

The modifications we made to the code include:
\begin{itemize}
\item change of the cubic interpolation process, going from (Mass, Age, Metallicity) to (EEP, Age, Metallicity) where EEPs are equivalent isochrone evolutionary points.\footnote{EEPs were introduced in Refs. \cite{Simpson70, Bertelli90, Bertelli94}. EEPs are a uniform basis which simplifies greatly  the interpolation among evolutionary tracks. Each phase of stellar evolution is represented by a given number of points, each point in one track has a comparable interpretation in another track.} EEPs are provided by  \texttt{isochrones}, we only modify  the interpolation interface, following the implementation of \texttt{isochrones}2.0,
\item implementation of   a standard magnitude correction to account for extinction in the selected filters according to the Fitzpatrick extinction curve (see e.g., Ref.~\cite{Fitz99}) in the selected (here HST $F_{606W}$ and $F_{814W}$) and V band filter,
\item interpolation on the [$\alpha$/Fe] parameter. 
\end{itemize}
 
The  set of fitted parameters for each GC are age, distance modulus, metallicity, [$\alpha$/Fe] and absorption. Note that there are different photometry files corresponding to different values of  metallicity [Fe/H] and Helium fraction, $Y$.\footnote{The Helium fraction $Y$  of a GC is not necessarily identical to the cosmological one. If Population III stars  have enriched the medium with Helium, it is the resulting Helium fraction that matters here. Hence, in principle there could be object by object (GC) variations of $Y$. }  These, however,  are not two fully independent quantities: both quantities are a function of the  stellar and (proto)-solar metal mass fraction, denoted by $Z$ and $Z_{\odot}$, respectively. Consequently, they are  highly correlated. We are interested in the Age-Metallicity relation, hence for our purposes we can use only one of them, the [Fe/H] fraction\footnote{The metallicity $Z$ is related to [Fe/H] fraction in the usual way:  [Fe/H] = 1.024 log(Z) + 1.739, see Ref.~\cite{Bertelli94}.} in our case, as the independent variable. We vary [$\alpha$/Fe]  independently of [Fe/H] and $Y$.

\section{Color-magnitude diagram-based likelihood for globular clusters}
\label{sec:calib}

As mentioned in Sec.~\ref{sec:intro}, the traditional Bayesian analysis of this kind of data sets attempts to model each star independently, which implies a significant computational cost due to the large number of parameters to explore. A common approach is to fit the initial mass of each of the $N_{\rm stars}$ stars in the color-magnitude diagram  as an independent parameter (along  all other stellar parameters). Then, the posterior is marginalized over all individual star parameters to infer the parameters describing the GC.

\begin{figure}
\centering
\includegraphics[width=0.7\textwidth]{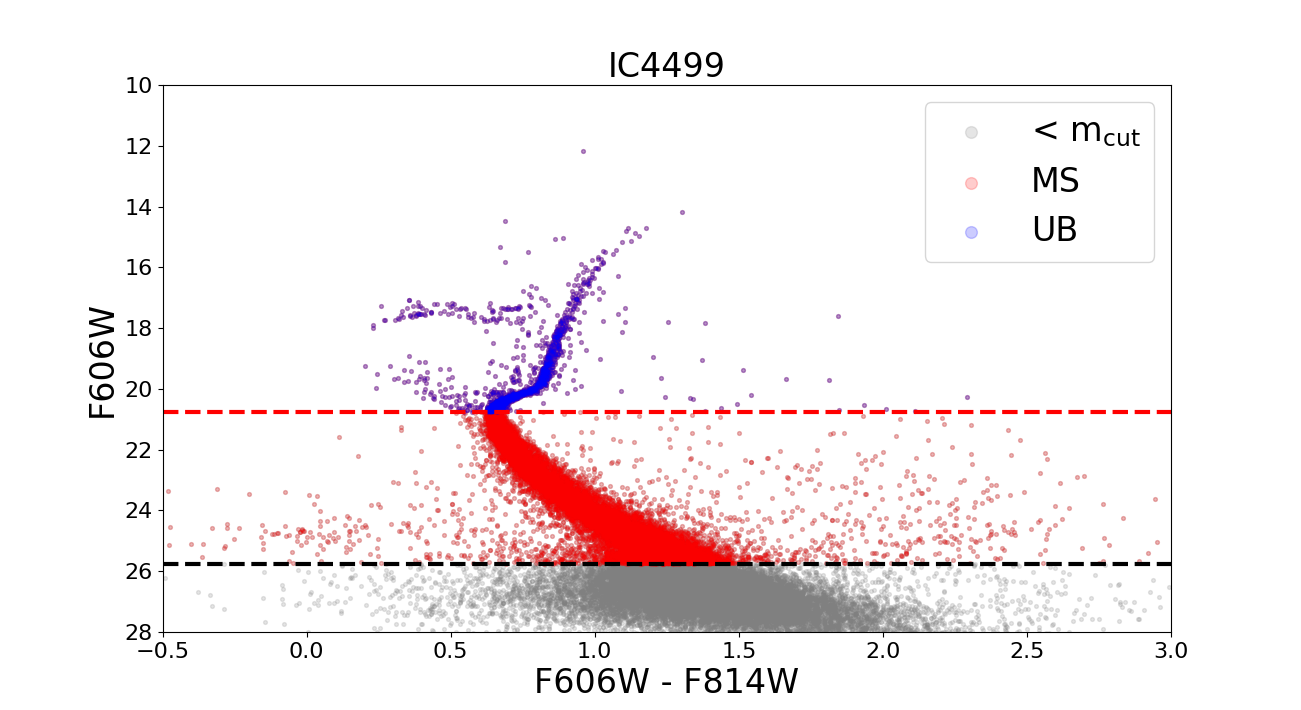} 
\includegraphics[width=0.7\textwidth]{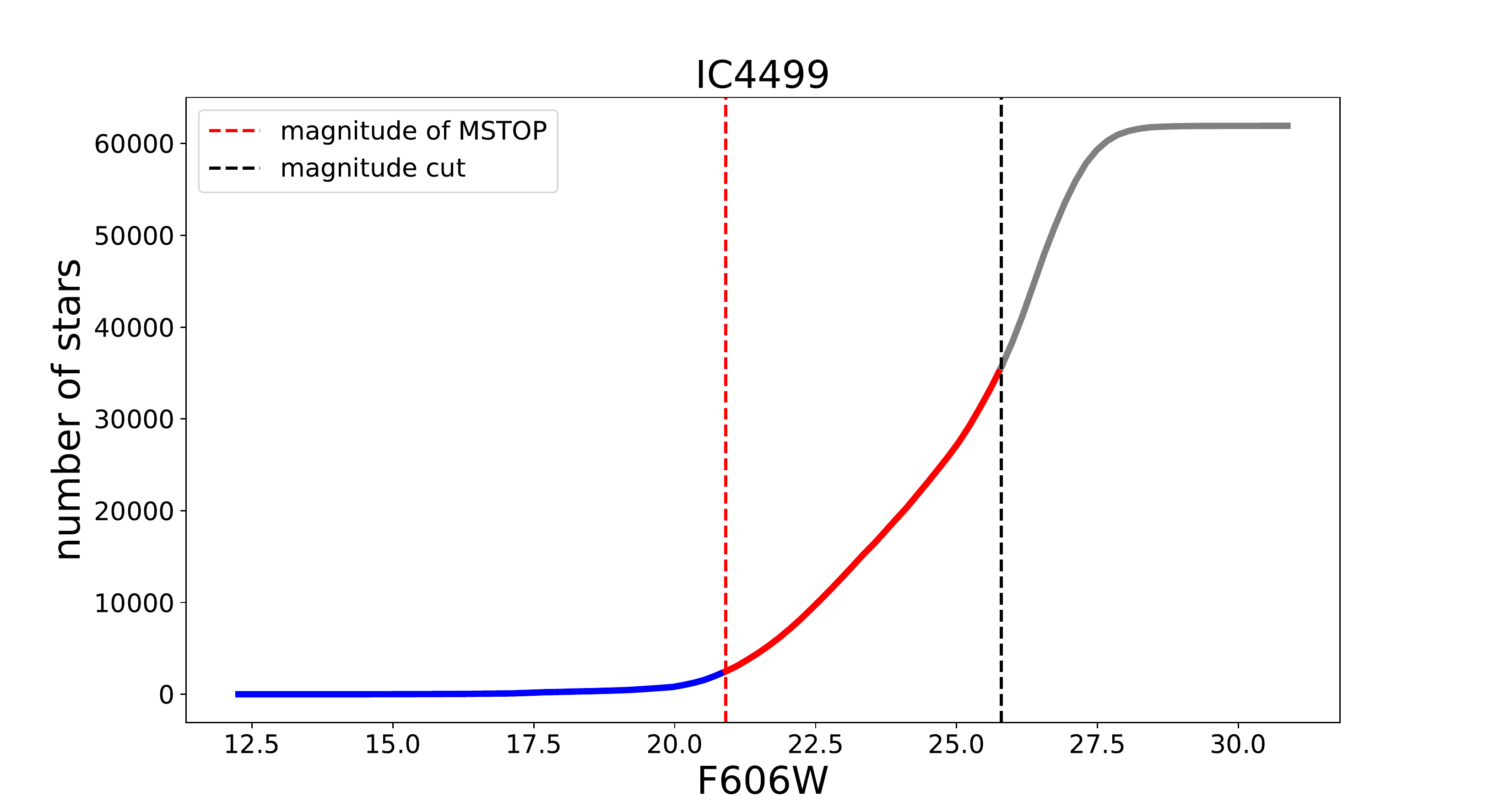} 
\caption{Top Panel: Illustration, for a typical GC (IC4499),  of  the  initial split of the ``functional''  magnitude interval in two parts (MS below the MSTOP and UB above the MSTOP). The red line corresponds to $m_{\rm MSTOP}$, and the black line marks $m_{\rm cut}$.  Points below $m_{\rm cut}$ do not add significant additional information,  but significantly slow down and complicate our analysis.  This is why they are  not considered here. Bottom panel: Cumulative distribution of stars  and adopted  magnitude cuts for the same cluster.}
\label{split}
\end{figure}

Here we attempt to reduce the high dimensionality of the parameter space using a different approach. While the  large number of stars can be a liability in terms of computational cost for traditional Bayesian approaches, we turn it to our advantage, especially in the most populated  part of the  color-magnitude diagram. For each isochrone of the stellar model, there are a number of  equivalent evolutionary points (EEPs) 
(see line 5 of Table \ref{tab:Table1}) associated with an initial stellar mass. Each  EEP has a  counterpart in every isochrone, making it possible to identify specific points in the color-magnitude diagram across different isochrones, e.g., the MSTOP. In other words,  the isochrone profile in the color-magnitude diagram is sampled by EEPs (which are ``universal"   across different  isochrones) obtained for different  adopted values of the parameters of interest. This is the reason why, as it is well known, the interpolation   between evolution tracks  is greatly simplified by interpolating instead  directly between EEPs.
 Since we are not interested in the initial mass of stars, we  do not model each star independently and exploit the benefits of the EEPs  working directly  with them, as provided  by  the relevant photometry files. This reduces the dimensionality of our analysis to just the five GC parameters described in the previous section.
 
We divide the  ``functional'' magnitude interval into two parts as illustrated in Figure~\ref{split}: the part below the MSTOP,  which we refer to as MS for  main sequence, and the part above, which we refer to as UB for upper branch.  The large spread of colors at low magnitudes introduces a lot of noise,  which  slows down significantly the convergence of our algorithm without adding,   in practice, any useful additional signal. For this reason, on top of the selection cuts described in Sec.~\ref{sec:data}, we apply a potentially more stringent upper magnitude cut. In practice, for the 68 clusters in our catalog we choose an upper cut magnitude value
\begin{equation}
m_{\rm cut} = {\rm min}(m_{\rm MSTOP} + 5\,, 26),
\label{mcut}
\end{equation}
where $m_{\rm MSTOP}$ is the magnitude corresponding to the MSTOP. In fact, for some GCs going 5 magnitudes below the MSTOP would cause to include noisy data. With this choice we limit the cut for those GCs to $m_{\rm cut} = 26$. Our findings are not sensitive to the details of this cut as long as the noisy,  dim part of the color-magnitude diagram is removed, and enough EEPs in the main sequence are retained, which is what we ensure here.

\subsection{Main sequence}

\begin{figure}
\centering
\includegraphics[scale=0.55]{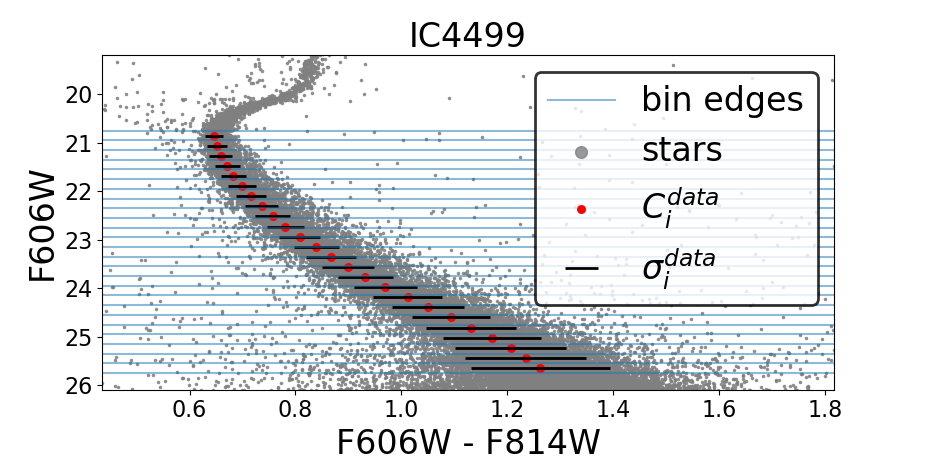} 
\caption{Binning of the main sequence, illustrated for the GC IC4499. The red dots and black lines represent  the central value and standard deviation of the color distribution in each bin, respectively.}
\label{fig:ms_binning}
\end{figure}

We proceed to  bin in magnitude the sample of stars belonging to the main sequence; these bins should be  thin enough so that the isochrone can be approximated as linear in each bin, yet with  number of stars per bin large enough  to satisfy the central limit theorem. Given the large number of stars in the MS (as illustrated in the bottom panel of Figure~\ref{split}), these two conditions are fulfilled for all GCs.  In practice,  we  use bins in  the $F_{606W}$ magnitude interval for the MS with constant width of 0.2 mag,  which yields  a maximum of 25 bins  and a minimum of  20 for the GCs in our catalog. The number of stars per bin is  proportional to the number of stars in the GC and ranges  from several hundreds to several thousands. 
 It is then justified to assume that the scatter in color of stars inside each magnitude bin follows a Gaussian distribution centered on the true underlying isochrone. This simplification (akin to a  coarse-graining in the color-magnitude diagram, and thus to a data-compression) alone allows us to decrease the effective size of the data set, and thus, compared to previous approaches, to reduce the number of model parameters for this part of the analysis: we have 5 parameters, and $N_{\rm bins}$ number of data points. The main peak of the distribution of  star positions along the color axis in each  bin, indexed by $i$,  should  be, and is, well approximated by a Gaussian distribution (see  Figure~\ref{fitms2} in Appendix~\ref{app:MScalib} for an illustration).  Bins where the distribution cannot be fit by a unimodal Gaussian -- a possible sign of multiple populations -- are removed from the analysis. This always happens at the faint end of the  main sequence (except for three clusters for which one to two bins are removed), even after the cut from Equation~\eqref{mcut}. For this reason we use instead the median of the distribution. It allows us to keep the maximum of 25 bins while taking into account the effect of the multiple population. The median value is almost identical to the Gaussian mean and larger error bars are a reasonable trade-off for outliers. More details are presented in Appendix \ref{app:MScalib}.  The color at bin center for each magnitude bin  $C^{\rm data}_{i}$ is defined by the median. Since the main sequence in the color-magnitude diagram is not perfectly vertical, we rescale the error by  $\sigma^{\rm data}_{i} \approx 1.253\sigma_{{\rm EEP},i} \times \cos (\phi_{i}) $ where $1.253\sigma_{{\rm EEP},i}$ is the standard error of the median and $\phi_{i}$ is the angle between the data orientation and the vertical axis inside bin $i$ as detailed in Appendix~\ref{app:MScalib} (in particular see  Figure~\ref{fitms1} in the Appendix). This correction is very small and always well below 4\%. Figure~\ref{fig:ms_binning} shows an example of this binning for  GC IC4499, along with the corresponding  $C^{\rm data}$ and $\sigma^{\rm data}$.

Assuming that bins are uncorrelated (which given the small observational errors in the star magnitudes is a fair assumption), the logarithm of the likelihood is defined as
\begin{equation}
\mathcal{L}_{MS} = \ln\: L = -\frac{1}{2} \sum_{i = 1}^{N_{\rm bins}} \left(\frac{C^{\rm data}_{i} - C^{\rm th}_i}{\sigma^{\rm data}_{i}}\right)^2
\label{eq: likelihood}
\end{equation}
where $C^{\rm th}_i$ is the theoretical isochrone color  interpolated at the center of bin $i$, and $N_{\rm bins}$ is the number of bins considered in the analysis (i.e., after removing the bins with bimodal color distributions).

\subsection{Upper branch}
\label{sec:ub}
In addition to the main sequence, we consider stars belonging to the Upper Branch (UB) i.e., stars brighter than the MSTOP. We bin the magnitude interval as we did for the main sequence. However, in this case, the number of stars is not large enough to support the central limit theorem for  small magnitude bins; in addition we expect that the measurement will be highly sensitive to outliers. Therefore, we cannot fit the color distribution to a Gaussian function as done for the MS. Instead, we apply these three prescriptions:
\begin{itemize}
\item  Since  \texttt{DSED} isochrones do not include stages beyond the tip of the red giant branch  --i.e., do not include EEPs belonging to the Horizontal branch  and the asymptotic giant branch--,  we mask out all the bins which correspond to stars (and  EEP) that do not belong to either the sub-giant branch or the red giants.

\item Since the estimation of the  mean  is easily contaminated by outliers, we  use the median  color instead in each bin  as an estimate for $C^{\rm data}_i$. In fact, we expect that the color errors follow a Gaussian distribution, and that  the outliers  are stars that  are  not part of the GC main sequence of upper branch (our target sample).  If we could select only stars that belong to our target sample, they would follow a Gaussian distribution.   In practice, using the median down weights the contribution of outliers on the estimate of the central value of the distribution. Therefore, it provides a good estimate of the mean value of the distribution of the target sample; here we {\it assume} that the resulting distribution matches the target distribution and can be assumed to be Gaussian. 
\item  We use the error of the median for normal distributions  $\sigma_{{\rm med},i} = 1.253 \sigma_{{\rm EEP},i}$, where $\sigma_{{\rm EEP},i}$ is the regular standard deviation in bin $i$.
\end{itemize} 

\begin{figure}
\centering
\includegraphics[scale=0.4]{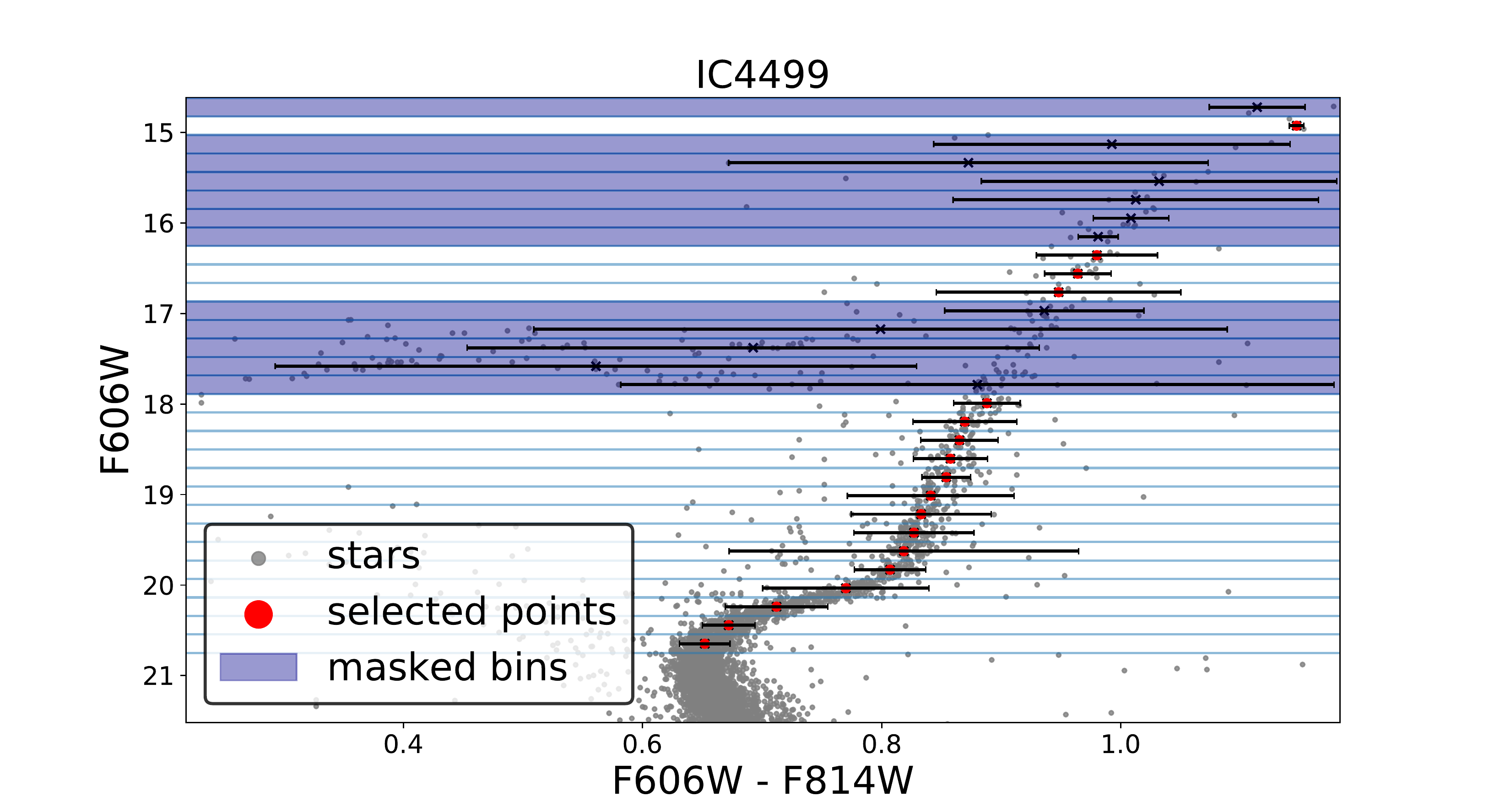} 
\caption{Binning of the upper branch for a representative GC IC4499.  The grey points are the stars, the horizontal blue lines show the adopted binning. The masked bins are  shaded. Each red point represents the median value at bin center.   The error bars correspond to $\sigma_{{\rm med},i}$.}
\label{fig:rgb}
\end{figure}

This is illustrated in   Figure \ref{fig:rgb}.   In this figure, for a representative GC, IC4999, the stars in the color-magnitude diagram  are shown as grey points, the excluded bins are shaded, the red points show the $C_i^{\rm data}$, and the error bars   show the $\sigma_{{\rm med},i}$. 

Finally the likelihood is also taken to be Gaussian as in Eq. \ref{eq: likelihood}, with the only differences of $C^{\rm data}_{i}$ being the median value at bin center, and $\sigma_{{\rm med},i}$ the associated error for bin $i$.  
We  are aware that  this choice of Gaussian likelihood is not as well  motivated  as for the MS.  Nevertheless we note here that other systematic  uncertainties (see section \ref{sec:syserr}) are likely larger than the one introduced by this approximation.

\subsection{Multiple populations and magnitude cut}
\label{sec:multiple_pop}
For the sake of simplicity in the  analysis, we assume that parameters such as age, metallicity and distance are common to all stars belonging to the GC. Nonetheless, GCs can be  more complex and  host distinct populations. Multiple populations in GCs is an active research area (see e.g., \cite{ReviewBastian18} for a review). It is important to note that multiple populations do not necessarily have different ages, they may have e.g., different element abundances. Moreover, the effects of multiple populations are minimized for the filters used to create the catalog ($F_{606W}$ and $F_{814W}$; see Ref.~\cite{ReviewBastian18} and references therein). When we apply our analysis to GCs  known to host multiple populations to quantify the effect that this might have in the inferred constraints, we find that having multiple populations introduces  an additional widening in the marginalized inferred age, as well as multiple peaks for the metallicities. GCs with multiple populations have a manifestly  multi-modal posterior distribution  where additional {\it local} maxima may  appear.   We find that the  magnitude cut $m_{\rm cut}$ (see Equation \ref{mcut}) we impose helps to reduce  the sensitivity to secondary populations, i.e., it suppresses the secondary local maxima, but leave the global maximum unaffected. 
This is because it is easier to see multiple population in the faint end of the MS; at brighter magnitudes, the two populations blend. Nevertheless, the  posterior distributions obtained for some GCs are still multi-modal. Masking out bins where the distribution is markedly multimodal further minimize this effect.  Any residual  multi-modality is blended with the main maximum and thus effectively contributes to growing the errors. The way we deal in practice with the  multi-modality of these secondary local maxima is developed further in Sec. \ref{sec:inference}.  

\section{Parameter inference}
\label{sec:inference}

We assume that the two parts (MS and UB) of the ``functional'' magnitude interval considered  are independent. The total log-likelihood, ${\cal L}=\ln L$, is then $\mathcal{L} = \mathcal{L}_{\rm MS} + \mathcal{L}_{\rm UB}$.

The parameters that we vary are: age, metallicity [Fe/H], absorption, distance and  $\alpha$ enhancement  [$\alpha$/Fe].
In order to ensure that we remain inside the interpolation domain of the stellar model, we use uniform priors corresponding to  the intersection of the parameter-space volumes of the stellar model (in our case this corresponds to the prior region of  \texttt{DSED} see Table  \ref{tab:Table1}). These are: [1,15] Gyr for the age, [-2.5,0.5] dex for metallicity, (0,3] for absorption, (0,$\infty$) for distance and  [-0.2,0.8] for [$\alpha$/Fe].

In addition, we adopt gaussian priors on the metallicity, distance modulus, absorption and [$\alpha$/Fe] as follows.
For the metallicity, $\alpha$ enhancement and distance  the priors are centered around  estimates from the literature for each globular cluster (see Ref.~\citep{Dotter2010}).  For 65 clusters the  extinction estimates are based on the two catalogs of Refs~\cite{Harris, Dutra}; however, for three globular clusters (NGC 6121, NGC 6144, NGC 6723) we  use instead  values from more recent literature (Refs~ \cite{ngc6121,ngc6144,ngc6723}  respectively)  since  the quality of the fit  and  the posterior were unacceptable when using  the catalogs estimates.  

We adopt   $\sigma_{\rm [Fe/H]} = 0.2$ dex for the width of the Gaussian priors for the metallicity, based on spectroscopic measurements,  corresponding to twice the typical errors reported in Ref.~\cite{Bolte+}).\footnote{In principle, this prior could be more stringent,  following  Ref.~\cite{Carretta}. However we decide not to do this here, and explore a wider range in metallicy.} The width adopted for the  distance modulus prior is  $\sigma_{\rm dm} = 0.5$ from Gaia/Hipparcos indirect distances, 2-3 times the typical errors reported in Ref.~\cite{Bolte+,OMalley}.  We assume a dispersion on the reddening  $\sigma_{E(B-V)} = 0.02$, in agreement with Ref.~\cite{OMalley}, which translates into Gaussian priors on absorption with $\sigma_{\rm abs} = 0.06$ following the Cardelli et al. \cite{Cardelli} relation. For  [$\alpha$/Fe] we adopt a prior of $\sigma_{\alpha} = 0.2$ which is equivalent to the sampling step of the DSED stellar grid.
 
 Unlike the priors on metallicity or distance which are conservative compared to recent literature, the prior on absorption needs to be restrictive to reduce the degeneracy between age and absorption. Even though it may appear narrow, one should bear in mind that  this parameter is usually kept fixed in other analyses in the literature.\footnote{We have also explored relaxing the metallicity prior by increasing the width of the gaussian by a factor few. We find that this more conservative choice does not affect the final  results of the inferred age ($t_{\rm GC}$, $t_{\rm U}$) as statistical errors remain below the systematic ones.} 
 
For some clusters, the posterior distribution is cut by the 15 Gyr age limit imposed by the grid of the stellar models, but  even in these cases the peak of the distribution is always well determined and the cut  happens at the $\sim 2 \sigma$ level, hence the effect on the results can be kept under control.

Given the nature of the problem (degeneracy between the age, distance and the metallicity), the nature of the data (possible presence of multiple populations), and the nature of the likelihood calibration (we fit, at the same time, the MS and the UB, where, in principle,  each might favor a different region of the parameter space and be affected by different degeneracies), we expect that  the posterior distribution might be multi-modal. In this case, 
the standard \texttt{emcee} sampler  may be inefficient.

Existing methodologies  to handle multi-modal distributions  include  slicing the parameter space and combining the results afterwards, or techniques like parallel-tempering Monte Carlo Markov chains  where the chains are run at different temperatures,  which makes it easier to the chains to communicate  and thus ``move" between peaks and low likelihood regions.
The  first approach  is expensive in terms of computational cost and we found  the second one  not efficient in our case.
 Parallel tempering MCMC will move the ``coldest" chains to a formal global maximum which  is however in a non-physical region of parameter space (ages $\gtrsim$ 15 Gyr and  very  low metallicities [Fe/H] $<$ -2.3 dex).  We explain this tendency as follows.  At high ages and low metallicities the  evolutionary tracks in the color magnitude diagram become very similar to each other (as shown in Figure~\ref{diff1} in Appendix~\ref{sec:sensitivity}). In other words, there is a lot of  prior  volume  to explore, and therefore the chains tend to spend a lot of time there.  This is an artifact of the  prior  probability distribution chosen.

One of the consequences of having multi-modal posterior distributions with several local maxima of the likelihood and one global maximum, and using the standard affine invariant \texttt{emcee} sampler, is a low acceptance fraction. This is especially significant if the modes are well separated, i.e.,  if the separation between modes is much larger than the width of the distribution around the maxima. Indeed, only a small fraction of MCMC steps, close to the  likelihoods peaks are accepted. One possibility  to bypass this technical difficulty may involve re-parametrization~\cite{KosowskyJim} or non-uniform priors, in addition to using stronger Gaussian priors on the metallicity.

We decided to stick to the standard \texttt{emcee} sampler and increased the number of chains to improve the number of accepted steps. We run 100 chains (or walkers for \texttt{emcee}) for 5000 steps (several times the autocorrelation length) with a burn-in phase of 500 steps. This set up returns a suitable and stable acceptance rate. For multimodal distribution, the initialization of the chains can be a important factor. We tested two configurations (a tiny Gaussian ball centered on estimates from the literature see Ref.~\citep{Dotter2010} and a uniform distribution with boundaries matching the uniform priors. Both gave consistent results and we kept the second configurations as it is more objective.
We have also made several convergence tests on a subset of clusters varying the number of walkers and increasing the steps of each of them (from 100 to 700 walkers for up
to 100,000 steps) and found that this does not change the results.

We report the error on the parameters as the highest posterior density interval (also sometimes referred to as minimum credible interval) at a given confidence level. Note that for non-symmetric distributions (such as those we have here) these errors are not necessarily symmetric.

\subsection{Systematic uncertainties}
\label{sec:syserr}
In our approach, all the parameters that describe the GCs (age, distance, metallicity,  [$\alpha$/Fe] and extinction) are determined directly from the data. While HST photometry does have some remaining systematic uncertainty, this is minute compared to the uncertainty associated with the theoretical stellar model (see below). We estimate the systematic uncertainties in the ages of GCs induced by the theoretical stellar model using the recipe in  Table~2 of Ref.~\cite{OMalley}. To our knowledge, this is the most rigorous approach among stellar model-building to estimate the systematic uncertainties using the ``known-unknowns". Inspection of Table~2 in Ref.~\cite{OMalley} shows that the main systematic uncertainty is due to the use of mixing length theory to model convection in the 1D stellar models.  The other dominant systematic uncertainty is related to reaction rates and opacities.\footnote{Rotation is another source of systematic uncertainty, as the rotation speed of stars in GCs is unknown. However, the main effect of rotation is to alter the depth of the convection zone. Given that we have explored a wide range of values of the mixing length parameters, the effect of rotation is effectively included in our systematic budget estimation.} Everything else is subdominant, thus the combined effect these two components captures well the extent of systematic errors.
 
Mixing-length theory\footnote{In Appendix~\ref{appendix:MLT} we give a brief description of mixing-length theory.} has two parameters: the mixing-length parameter (i.e., roughly how much the convection cells travel before they break up), and the overshoot parameter (how much the convective cell travels beyond the equilibrium condition). Of these two, the second one is unimportant for low-mass stars such as those in GCs. These two parameters dominate the uncertainty in stellar model building; the uncertainties in nuclear reactions are at the \% level. 

In principle, changes in the mixing length do not alter the lifetime of the star, see discussion  in page 725, of Ref.~\cite{Jimenez95}. The effect on the inferred age arises  from degeneracies with metallicity. In this work the metallicity is strongly constrained so that, in principle, the effect of mixing length uncertainties could be reduced significantly. 

In fact, the mixing length parameter is usually calibrated from fits to the Sun, but astro-seismology from other stars at different evolutionary stages indicates a spread of values between 1.0 and 1.7.  Thus, the results from observations of the Sun are  extrapolated to stars belonging to GCs, but adopting the full spread of   mixing length parameter values to  quantify  the systematic uncertainties.
  However, a better estimation of systematic uncertainties related with the mixing length parameter is possible. As shown in Ref.~\cite{JimenezGC96}, not only the morphology of the red giant branch can be used to constrain the value of the mixing length, but also all the GCs analysed in Ref.~\cite{JimenezGC96}, had the same value for the mixing length and  showed  no  star-to-star variation of the mixing length parameter.  Therefore,  the morphology of the red giant branch is sufficient to constrain the mixing length, once the metallicity is constrained, without the need to rely on the solar calibration. Thus, potentially, for  the present study, as  the metallicity can be constrained from the lower main sequence as well as the sub-giant branch (see  Figure~\ref{diff1}), the upper giant branch could  be used to  determine the value of the mixing length as done in Ref.~\cite{JimenezGC96}. This approach would require adding the  mixing length parameter as an extra free parameter in our analysis;  we leave this for future work. 
  
Here instead we prefer to be conservative and use the full range for the  mixing length considered in Ref.~\cite{OMalley} (i.e., between 1.0 and 1.7), which is conservative because the study of  Ref.~\cite{JimenezGC96} showed that fits to the position of the red giant branch with known metallicity indicate no  significant spread in mixing length parameter. These fits recover a value of 1.6, well in agreement with  results from calibration to the Sun. To estimate the error in ages due to  mixing length  variations over the full conservative  interval,  we use the stellar models of Ref.~\cite{Jimenez95},  and in particular the fitting formulas therein. This yields  a 0.3 Gyr  age uncertainty.
 
In addition to this we add an extra 0.2 Gyr  to account for uncertainties in reaction rates and opacities, as from  Table~2 of  Ref.~\cite{OMalley}. In total, we have a 0.5 Gyr uncertainty budget due to systematic effects in stellar modeling.

Note that in the standard MSTOP approach, another systematic uncertainty to account for would be the value of [$\alpha$/Fe], which in general is not known and is assumed to be between $0.2-0.4$. However, in our  approach,  this is not the case as this is a parameter of the model: its value  is directly constrained by the analysis and its uncertainty is therefore already included in our marginalized errors.

\section{Results}
\label{sec:results}

\begin{figure}
 \begin{centering}
\includegraphics[width=0.49\columnwidth]{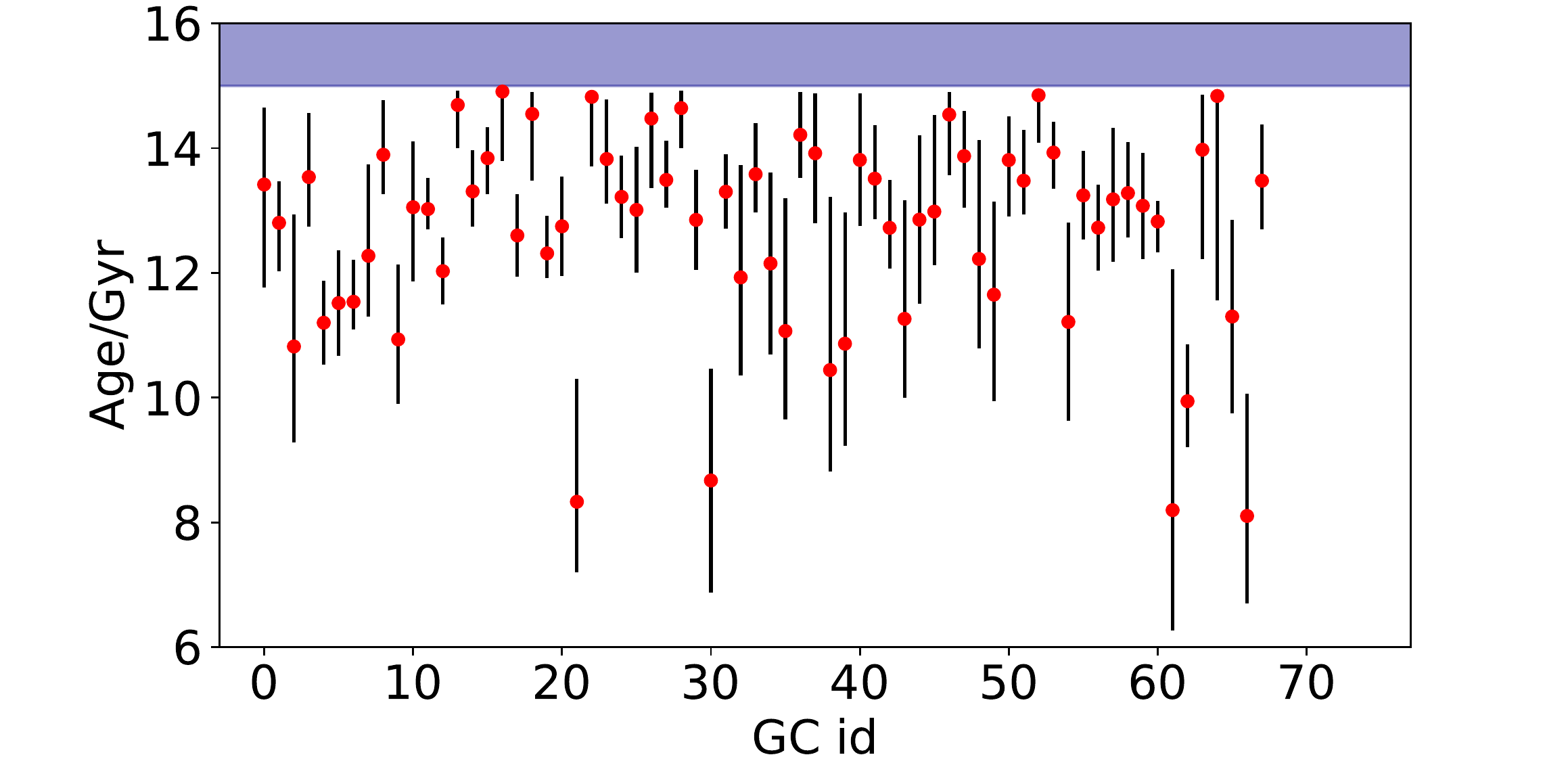}
\includegraphics[width=0.49\columnwidth]{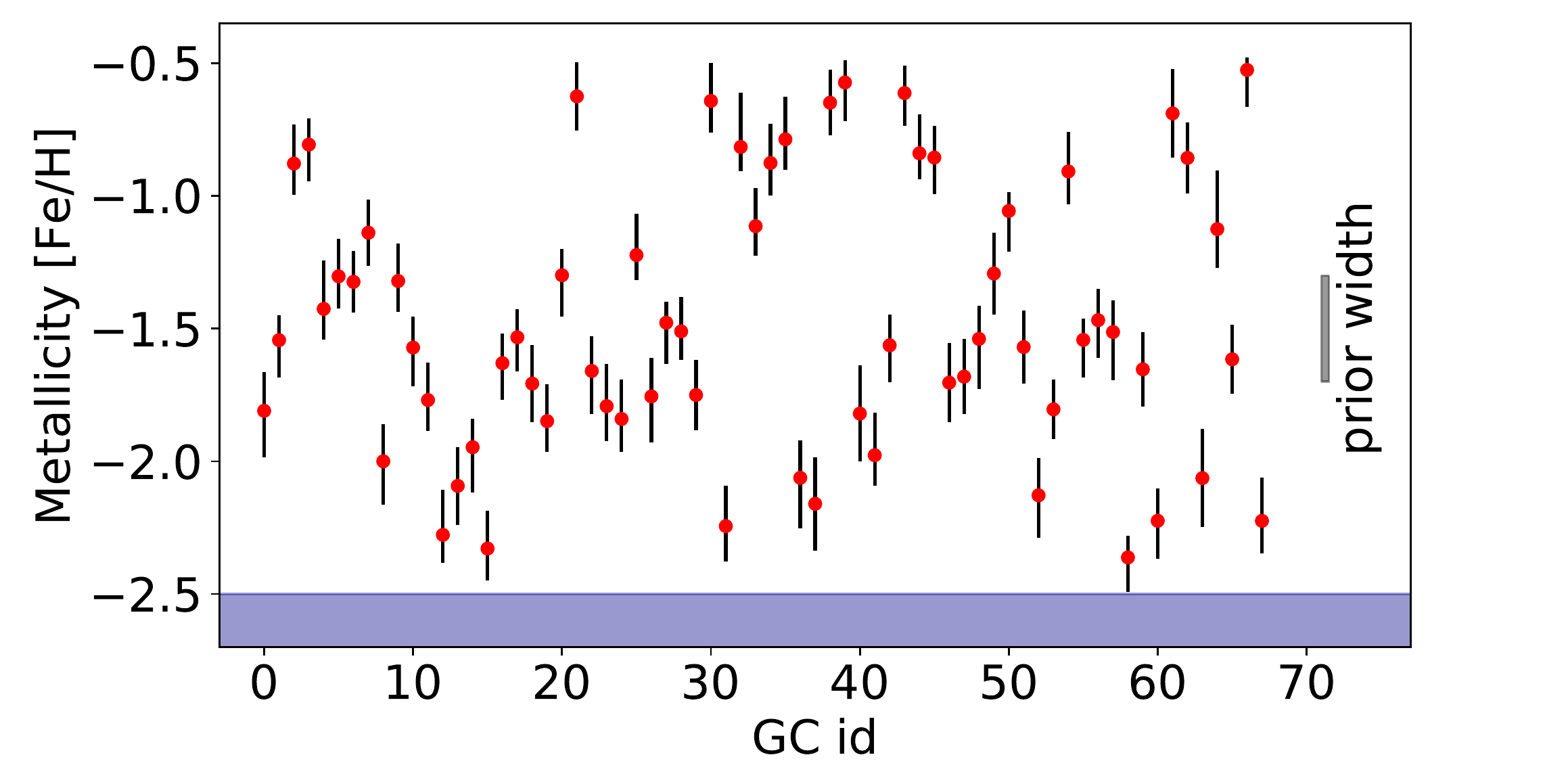}
\includegraphics[width=0.49\columnwidth]{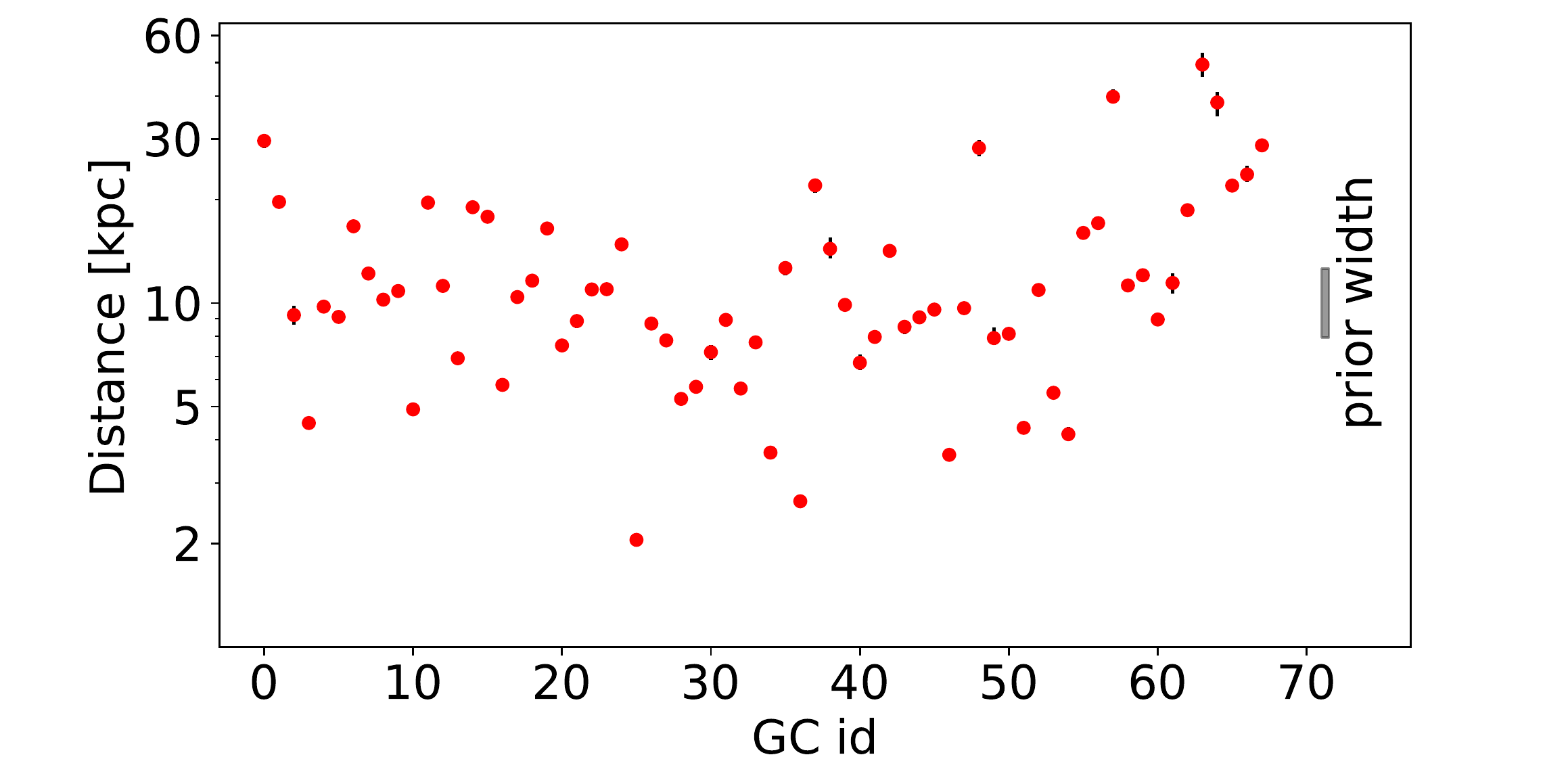}
\includegraphics[width=0.49\columnwidth]{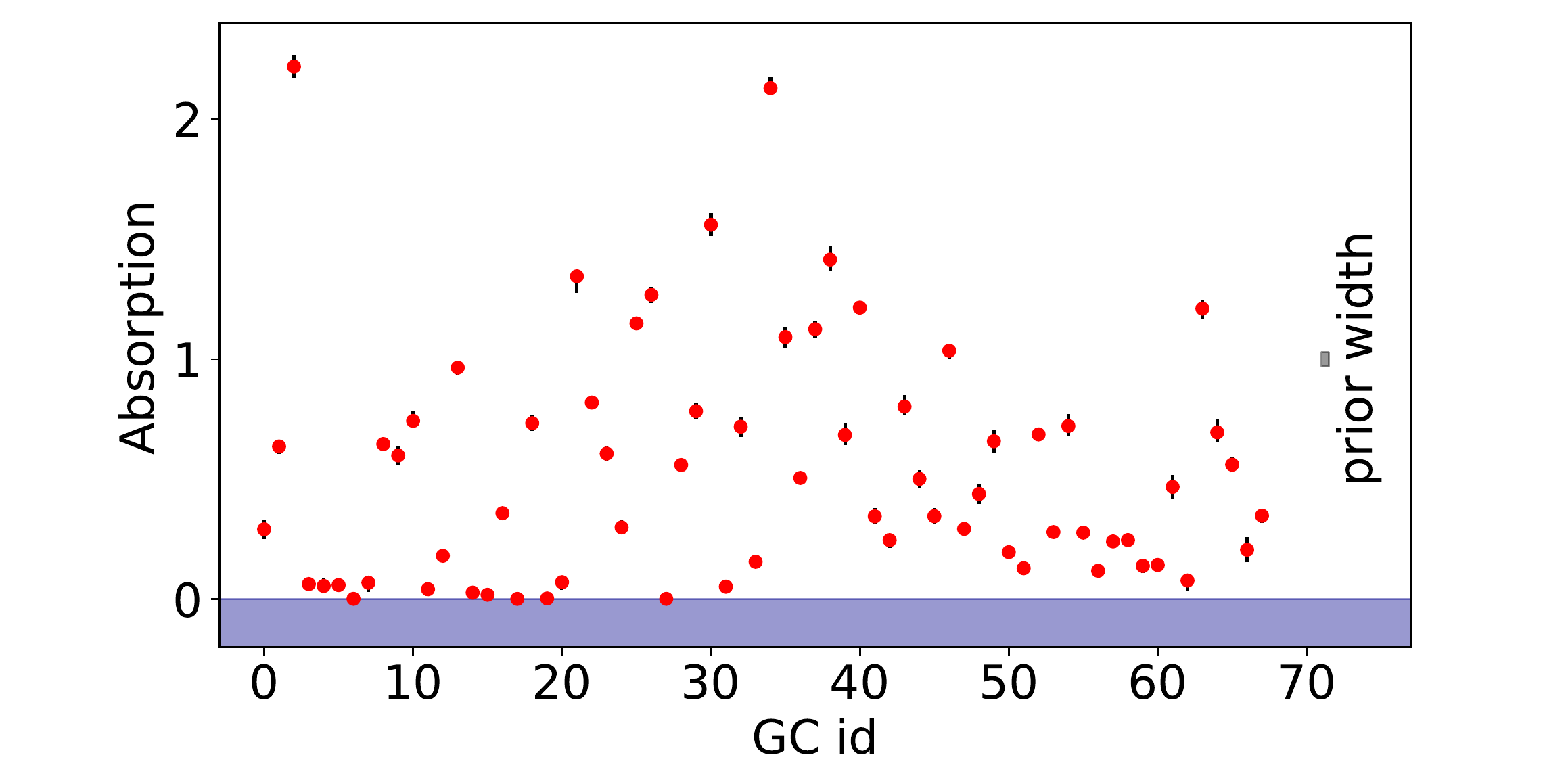}
\includegraphics[width=0.49\columnwidth]{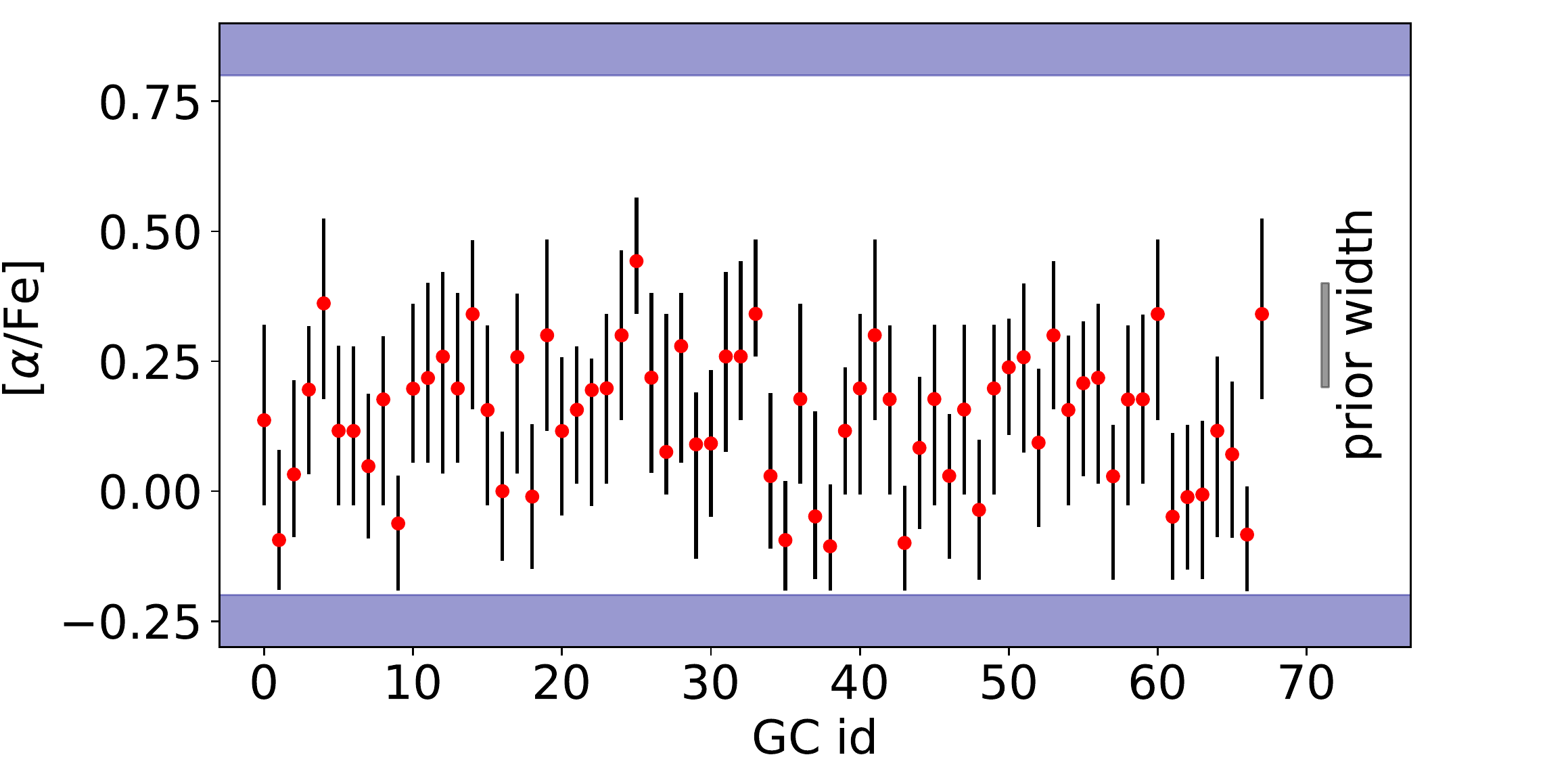}
\caption{$68\%$ confidence level marginalised constraints for the five parameters of interest
  for each of the GC in the sample (CG id, in the $x$-axis, corresponds to the ordering of Table \ref{tab:my-table}). The shaded blue regions represent  boundaries of the uniform prior. There are additional gaussian priors of $\sigma_{[Fe/H]} = 0.2$ dex for metallicity, $\sigma_{\rm dm}$ = 0.5 on the distance modulus, $\sigma_{[\alpha]} = 0.2$ for alpha enhancement and $\sigma_{\rm abs}=0.06$ in the absorption centered around values from the literature (see text for details).}
\label{fig:GC1}
\end{centering}
\end{figure}

We apply the methodology presented in previous sections to our catalog of 68 GCs.  Two-dimensional marginalized posteriors for all pairs of parameters  can be found for 
a representative GC in Appendix~\ref{app:fits}.
Figure~\ref{fig:GC1} shows our main results (see also Appendix~\ref{app:GCtable-params} and Tables~\ref{tab:my-table} and \ref{tab:my-table2}). We present marginalized constraints on the absolute age, metallicity, distance,  absorption  and [$\alpha$/Fe] of each GC assuming the \texttt{DSED} model. The $x$-axis in each panel shows the cluster id  following the same order as  in table~\ref{tab:my-table}.  The gray  horizontal areas show the hard priors  imposed by the stellar models domain in parameter space  and the gray vertical band (when reported) illustrates the width the gaussian prior adopted (see Sec.~\ref{sec:inference}).  
We find no correlation between age   and  distances,  absorption  or [$\alpha$/Fe].  In particular the absorption  values  are low  and the distribution  presents a scatter that is not correlated with the age. On an individual cluster-basis the constraints on  [$\alpha$/Fe] are very weak, however values of   [$\alpha$/Fe]$>0.6$ are typically disfavored. A comparison with Dotter et al. \citep{Dotter2010}  spectroscopic measurements can be found in  Appendix~\ref{app:fits}. 

\begin{figure}
 \begin{centering}
\includegraphics[width=1.\columnwidth]{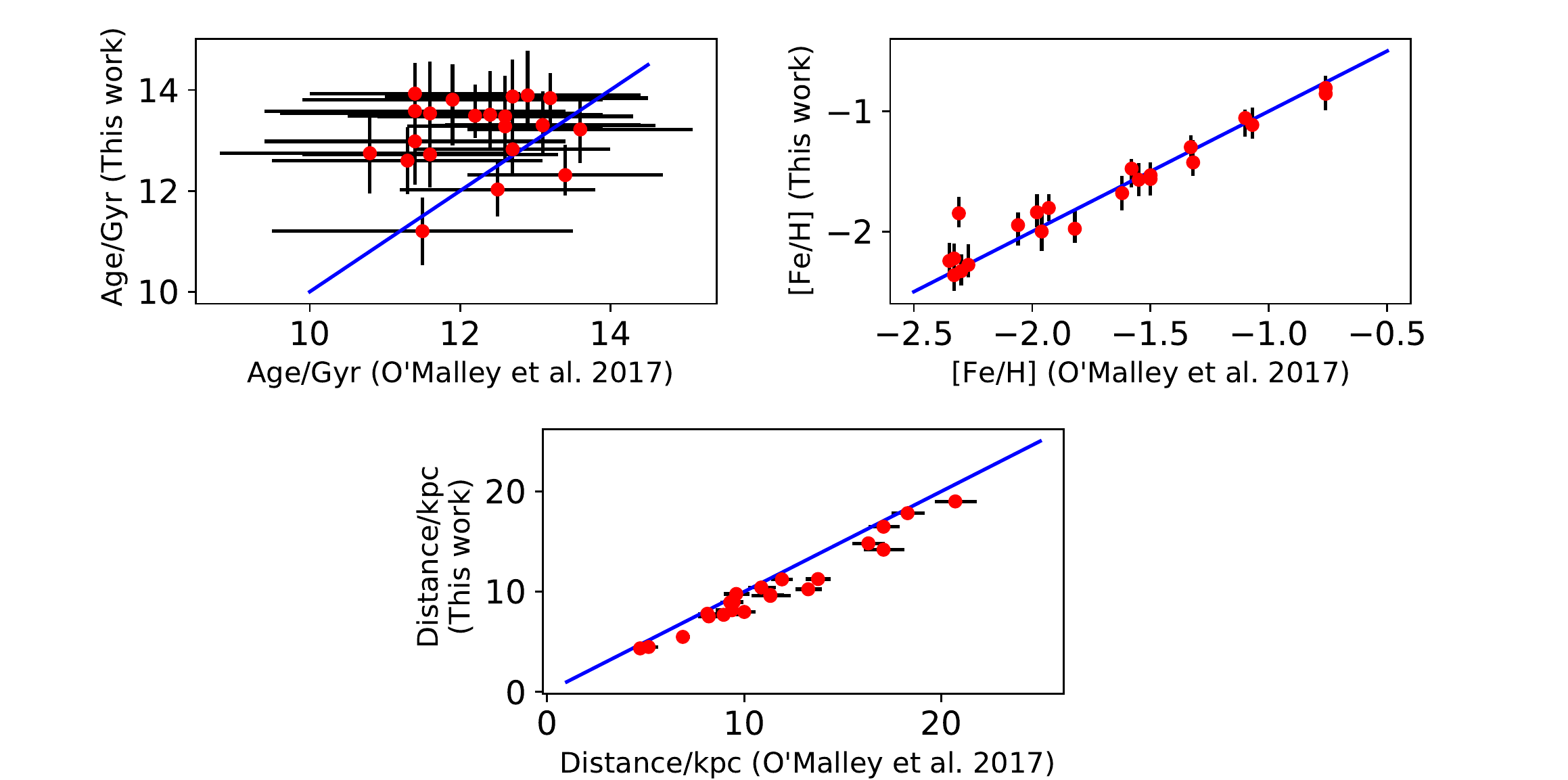}
\caption{Direct comparison between our marginalized constraints on the age, distance and metallicity of GCs with results from Ref.~\cite{OMalley} for the 22 GCs in common. The blue lines indicate the identity. We  plot uncertainty bars for both determinations when available.  There is excellent agreement for the metallicity determination and  reasonable agreement for the distance determinations, although our distances (with error bars so small that are behind the red dots) are on average somewhat shorter than those of Ref.~\cite{OMalley} by about  200 pc.  The age agreement is within the uncertainties, but our ages are slightly older on average. See text for more details.}  
\label{fig:GC4}
\end{centering}
\end{figure}
 
In Figure~\ref{fig:GC4} we compare our inferred  constraints with the findings of  Ref.~\cite{OMalley} for the 22 GCs in common. It is interesting to note the good agreement obtained for the metallicity estimates of [Fe/H]. Our distances, using  information from the color-magnitude diagram and only very weak priors,  are in reasonable agreement with those obtained  Ref.~\cite{OMalley}, which rely on  external information (GAIA parallaxes and accurate distance to nearby dwarf stars). However, we find a small shift as our determination of distances is $\sim 200$ pc smaller on average. This small discrepancy arises because the analysis in Ref.~\cite{OMalley} assumes a fixed extinction value, while we treat extinction as a free parameter to be constrained by the data and marginalized over.  For the ages determination the agreement  is  within 68\% confidence level uncertainties. From the first panel of  Figure~\ref{fig:GC4} it is possible to appreciate  that the errors from this study are smaller that those of Ref.~\cite{OMalley} even when Ref.~\cite{OMalley} uses  additional external information, not used here. This illustrates the advantage of considering regions of the color-magnitude diagram beyond the main sequence.

The use of the full color-magnitude diagram, along with the adoption of the priors motivated in sec.~\ref{sec:inference},  enables us to  break the age--distance--metallicity degeneracy. In particular, the breaking of the   age-metallicity degeneracy  is visualized in Appendix \ref{sec:sensitivity} where we show how the isochrones and the color magnitude diagram change in response to  variations  of these parameters.

\subsection{The age of the oldest GCs}
\label{sec:ageGC}
\begin{figure}
 \begin{centering}
\includegraphics[width=\columnwidth, height=0.4\textheight ]{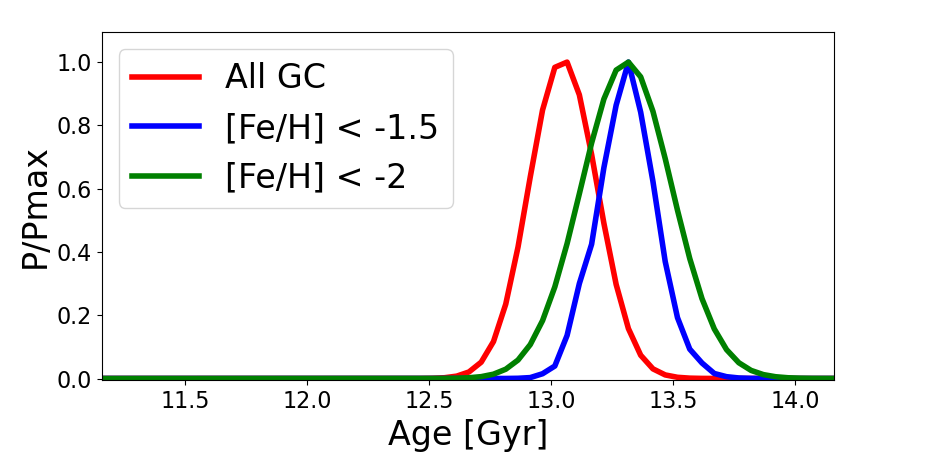}
\caption{ Age distribution for globular clusters with different metallicity cuts ([Fe/H] $< 2$ (dot-dashed); [Fe/H] $< 1.5$ (solid); no cut (dashed)) . The behavior is consistent with the expected age-metallicity relation. We only display the statistical uncertainty. An additional uncertainty of $0.5$ Gyr at 68\% confidence level needs to be added to account for the systematic uncertainty.}
\label{fig:GCage}
\end{centering}
\end{figure}

On average, the oldest GCs are those expected to  be more metal poor. Here we consider two metallicity cuts as a way to select the oldest GCs: [Fe/H]$<-2$ as adopted in Ref.~\cite{JimGC} -- leaving 11 clusters -- and  [Fe/H]$<-1.5$ -- leaving 38 clusters-- .    We estimate the age distribution  $t_{\rm GC}$ for these two samples by multiplying the individual bayesian posteriors (see Fig.~\ref{fig:GCage}).

For [Fe/H]$<-2$ this yields $t_{\rm GC}= 13.32 ^{+0.15}_{-0.20} {\rm (stat.)} \pm 0.5 {\rm (sys.)}$, while 
for  [Fe/H]$<-1.5$ we obtain $t_{\rm GC}= 13.32 \pm 0.1  {\rm (stat.)}   \pm 0.5 {\rm (sys.)}$. The first uncertainty is the statistical uncertainty while the second uncertainty is the systematic one, as calculated in Sec. \ref{sec:syserr}. The results for the two cuts are very  consistent; as expected, the additional 27 clusters in the   [Fe/H]$<-1.5$ cut reduce the statistical error significantly; here  we therefore adopt the  [Fe/H]$< - 1.5$ cut due to the increased statistical power.

\begin{figure}
\begin{centering}
\includegraphics[width=.8\columnwidth]{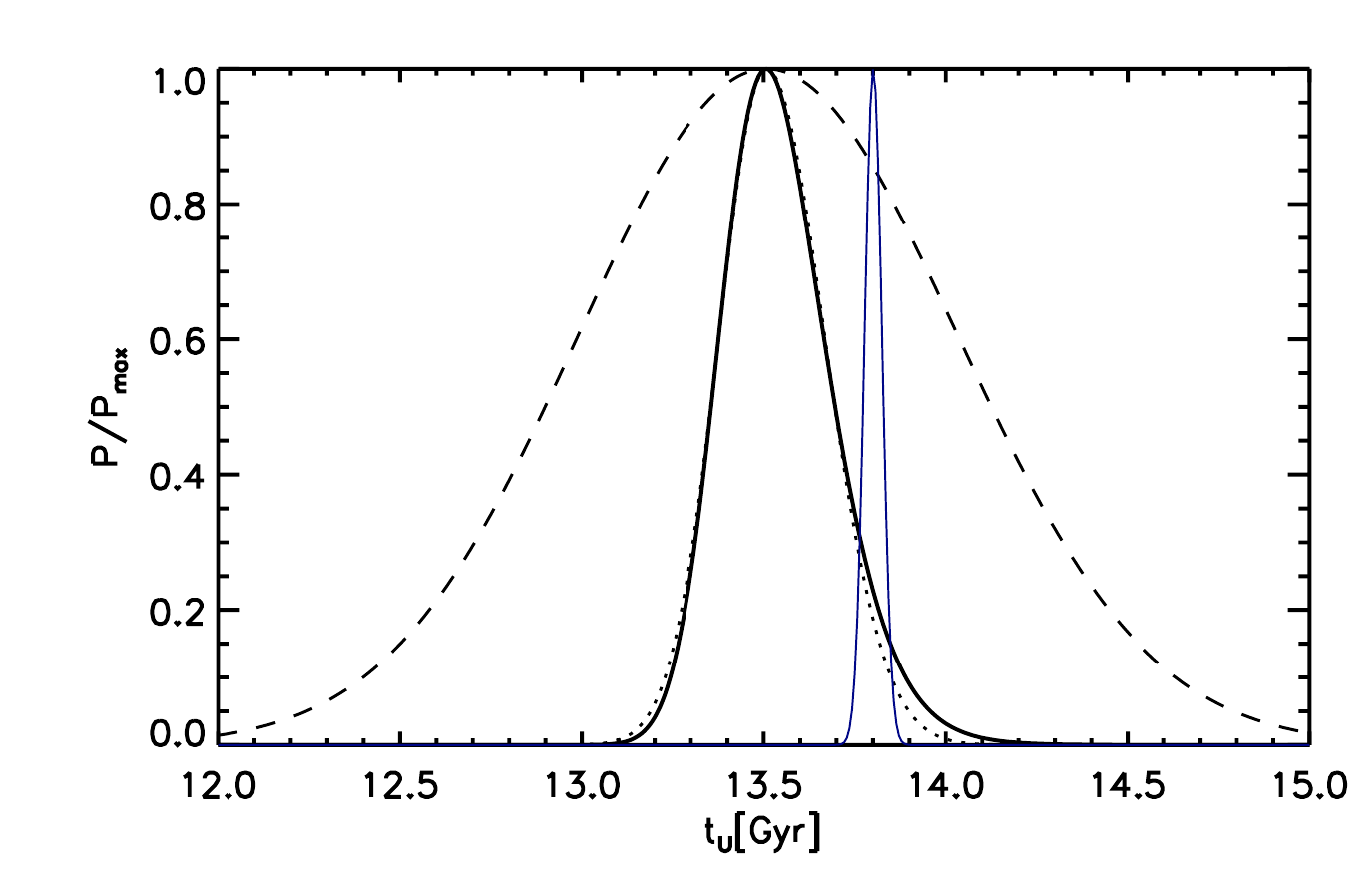}
\caption{Estimate of the age of the Universe from the age of the oldest globular clusters (solid  thick black line) including systematic uncertainties (dashed line) added in quadrature to a gaussian fit (with asymmetric variances) of the statistical distribution (dotted line). The thin blue line shows the  Planck 2018 posterior for the age of the Universe.}  
\label{fig:GCUniverse}
\end{centering}
\end{figure}

\subsection{From globular cluster ages to the age of the Universe}
\label{sec:ageuni}

The age of the oldest stars sets a lower limit for the age of the Universe. These stars and the oldest GCs formed at a redshift $z_{\rm f}$. Hence, it is possible to estimate the age $t_{\rm U}$ of the Universe  from the age $t_{\rm GC}$ of the oldest GCs adding  a formation time  $\Delta t$, corresponding to the look back time at $z_{\rm f}$. 

As argued in Ref.~\cite{JimGC},  it is possible to estimate  the probability distribution of $\Delta t$ by considering that the first galaxies are found at $z\sim 11$  and a significant number of galaxies are found at $z > 8$. Many of these galaxies contain stellar populations that indicate that star formation started at $z \sim 15-40 $~\cite{2018Natur.557..392H,2020ApJ...888..124S,2019MNRAS.489.3827B}; $z_{\rm f}$ is thus assumed to be $z_{\rm f}\ge11$. Both theoretically \citep{Padoan_Jimenez1997,Trenti2015,Choksi2018,Reina_Campos2019,Kruijssen2019}
and observationally \citep{Forbes2015} GC seem to form at $z_f >10$. On the other hand, GCs could not have formed before the start of reionization which is estimated to happen around $z_{\rm f, max}\sim 30$.  Ref.~\cite{JimGC} includes a computation of the probability distribution of  $\Delta t$ marginalizing over $H_0$, $\Omega_{m,0}$ and $z_{\rm f}$, with $z_{\rm f}$ varying between $z_{\rm f, min} = 11$ and $z_{\rm f, max}$. The resulting distribution depends very weakly on cosmology for reasonable values of the cosmological parameters, and very weakly on the choice of  $z_{\rm f, max}$ provided $z_{\rm f, max}> 20$. Here we estimate the full probability distribution of $t_{\rm U}=t_{\rm GC}+\Delta_t$ by performing a convolution of the posterior  probability distribution for $t_{\rm GC}$  as provided in \S~\ref{sec:ageGC} and the probability distribution of $\Delta_t$ from Ref.~\cite{JimGC} for which we provide a fitting formula in appendix \ref{appendix:fitDt}.

We find $t_{\rm U}=13.5^{+0.16}_{-0.14} {\rm (stat.)} \pm 0.5 ({\rm sys.})$ at 68\% confidence level. The resulting posterior distribution for the age of the Universe $t_{\rm U}$ is presented in Figure \ref{fig:GCUniverse}. The solid black line is the result including only statistical errors, the dashed line is obtained by  fitting this distribution with  two gaussians with the same maximum but independent variances for the two sides (dotted line), and then adding the systematic error in quadrature (dashed line). For reference the blue thin line shows the constraint inferred from CMB observations from Planck,  assuming the $\Lambda$CDM model ~\cite{Planck18}.

\section{Summary and Conclusions}
\label{sec:summary}

Resolved stellar populations of GCs provide an excellent data set to constrain the age of the Universe, which in turn is a key parameter in cosmology governing the background expansion history of the Universe. Since the mid 90's, estimates of the ages of GCs have been in the range $12-14$ Gyr consistently (see e.g. Ref.~\cite{JimenezGC96}). With current improvements in observational data and stellar modeling, it is possible to decrease the uncertainty on the ages by a factor 4. Given the high-quality of data obtained by HST and the improvement in the accuracy of stellar models, we have attempted  to estimate the physical parameters of GCs including their age, using as many  features as possible in their color-magnitude diagrams. 

It is well known that the MSTOP is very sensitive to the GC's age; however, it is also sensitive to distance, metallicity, and other parameters, due to significant degeneracies in parameter space. However, degeneracies can be in large part lifted  if other features of the color-magnitude diagram are exploited (see Appendix~\ref{sec:sensitivity}). 
 In this paper, we have analyzed a sample of 68 ACS/HST globular clusters using most of the  information in the color-magnitude diagram: specifically, the main sequence and red giant branch. We have developed a Bayesian approach to perform an analysis of each GC, varying simultaneously their age, distance, metallicity, [$\alpha$/Fe] and reddening  adopting physically-motivated priors based on independent measurements of distances, metallicities and extinctions found in recent literature. Our obtained posteriors yield constraints that  are fully compatible with previous, and independent, values in the literature. 
  
The average age of the oldest (and most metal poor)  GCs is  $t_{\rm GC}=13.32 \pm 0.1 {\rm (stat.)} \pm 0.5 {\rm (sys.)}$  Gyr.  The systematic errors are due to  theoretical stellar model  uncertainties and in particular uncertainties in  the mixing length, reaction rates and opacities. Systematic errors  are now  bigger than  the statistical error,  once constraints from several objects are combined. Hence, to make  further progress,   uncertainties in stellar model-building should be addressed.  
 
 This determination  can be used to estimate the Universe absolute age by taking into account the look back time at the likely redshift of formation of these objects. We find  the age of the Universe as determined from stellar objects to be $t_{\rm U}=13.5^{+0.16}_{-0.14} {\rm (stat.)} \pm 0.5 ({\rm sys.})$ at 68\% confidence level. The statistical error is 1.2\%; the error budget is dominated by systematic uncertainties on the stellar modeling.
 The prospect of determining the age of the Universe with an accuracy  competitive with current cosmology standards,  may serve to motivate  an effort to reduce uncertainties in  stellar-model building. This will be addressed in future work.
 The statistical uncertainty in $t_U$ is now sufficiently small  to warrant comparison to the CMB model-dependent inferred age, which is one of the most accurately quantities inferred from the CMB~\cite{Knox}. Thus comparing the CMB derived value to independent astrophysical estimates can yield precious insights into possible new physics, or support the $\Lambda$CDM model. Our determined value of $t_{\rm U}$ is fully compatible with the inferred value from the Planck mission observations assuming the $\Lambda$CDM model.

\acknowledgments

We  thank the stellar modelers for making their stellar models publicly available. We thank the anonymous referee for an useful and constructive report. We also thank David Nataf for very useful feedback. This work is supported by  MINECO grant PGC2018-098866-B-I00 FEDER, UE.  JLB is supported by the Allan C. and Dorothy H. Davis Fellowship.
JLB has been supported by the Spanish MINECO under grant BES-2015-071307, co-funded by the ESF, during part of the development of this project.  LV acknowledges support by European Union's Horizon 2020 research and innovation program ERC (BePreSySe, grant agreement 725327).
The work of BDW is supported by the Labex ILP (reference ANR-10-LABX-63) part of the Idex SUPER,  received financial state aid managed by the Agence Nationale de la Recherche, as part of the programme Investissements d'avenir under the reference ANR-11-IDEX-0004-02; and by the ANR BIG4 project, grant ANR-16-CE23-0002 of the French Agence Nationale de la Recherche. 
The Center for Computational Astrophysics is supported by the Simons Foundation. 


\begin{thebibliography}{10}

\bibitem{Catelan}
M.~{Catelan}, \emph{{The ages of (the oldest) stars}},  in \emph{Rediscovering
  Our Galaxy}, C.~{Chiappini}, I.~{Minchev}, E.~{Starkenburg} and
  M.~{Valentini}, eds., vol.~334 of \emph{IAU Symposium}, pp.~11--20, Aug.,
  2018, \href{https://doi.org/10.1017/S1743921318000868}{DOI}
  [\href{https://arxiv.org/abs/1709.08656}{{\ttfamily 1709.08656}}].

\bibitem{Soderblom}
D.~R. {Soderblom}, \emph{{The Ages of Stars}},
  \href{https://doi.org/10.1146/annurev-astro-081309-130806}{\emph{ARAA}
  {\bfseries 48} (2010) 581} [\href{https://arxiv.org/abs/1003.6074}{{\ttfamily
  1003.6074}}].

\bibitem{Bolte+}
D.~A. {Vandenberg}, M.~{Bolte} and P.~B. {Stetson}, \emph{{The Age of the
  Galactic Globular Cluster System}},
  \href{https://doi.org/10.1146/annurev.astro.34.1.461}{\emph{ARAA} {\bfseries
  34} (1996) 461}.

\bibitem{OMalley}
E.~M. {O'Malley}, C.~{Gilligan} and B.~{Chaboyer}, \emph{{Absolute Ages and
  Distances of 22 GCs Using Monte Carlo Main-sequence Fitting}},
  \href{https://doi.org/10.3847/1538-4357/aa6574}{\emph{ApJ} {\bfseries 838}
  (2017) 162} [\href{https://arxiv.org/abs/1703.01915}{{\ttfamily
  1703.01915}}].

\bibitem{Hoyle}
C.~B. {Haselgrove} and F.~{Hoyle}, \emph{{A preliminary determination of the
  age of type II stars}},
  \href{https://doi.org/10.1093/mnras/116.5.527}{\emph{MNRAS} {\bfseries 116}
  (1956) 527}.

\bibitem{Sandage}
A.~R. {Sandage} and M.~{Schwarzschild}, \emph{{Inhomogeneous Stellar Models.
  II. Models with Exhausted Cores in Gravitational Contraction.}},
  \href{https://doi.org/10.1086/145638}{\emph{ApJ} {\bfseries 116} (1952) 463}.

\bibitem{JimenezPadoanLF}
R.~{Jimenez} and P.~{Padoan}, \emph{{A New Self-consistency Check on the Ages
  of Globular Clusters}}, \href{https://doi.org/10.1086/310053}{\emph{ApJL}
  {\bfseries 463} (1996) L17}.

\bibitem{PadoanJimenezLF}
P.~{Padoan} and R.~{Jimenez}, \emph{{Ages of Globular Clusters: Breaking the
  Age-Distance Degeneracy with the Luminosity Function}},
  \href{https://doi.org/10.1086/303582}{\emph{ApJ} {\bfseries 475} (1997) 580}
  [\href{https://arxiv.org/abs/astro-ph/9603060}{{\ttfamily
  astro-ph/9603060}}].

\bibitem{JimenezPadoanGC}
R.~{Jimenez} and P.~{Padoan}, \emph{{The Ages and Distances of Globular
  Clusters with the Luminosity Function Method: The Case of M5 and M55}},
  \href{https://doi.org/10.1086/305593}{\emph{ApJ} {\bfseries 498} (1998) 704}
  [\href{https://arxiv.org/abs/astro-ph/9701141}{{\ttfamily
  astro-ph/9701141}}].

\bibitem{JimenezGC96}
R.~{Jimenez}, P.~{Thejll}, U.~G. {Jorgensen}, J.~{MacDonald} and B.~{Pagel},
  \emph{{Ages of globular clusters: a new approach}},
  \href{https://doi.org/10.1093/mnras/282.3.926}{\emph{MNRAS} {\bfseries 282}
  (1996) 926} [\href{https://arxiv.org/abs/astro-ph/9602132}{{\ttfamily
  astro-ph/9602132}}].

\bibitem{BayesianGC}
R.~{Wagner-Kaiser}, A.~{Sarajedini}, T.~{von Hippel}, D.~C. {Stenning}, D.~A.
  {van Dyk}, E.~{Jeffery} et~al., \emph{{The ACS survey of Galactic globular
  clusters - XIV. Bayesian single-population analysis of 69 globular
  clusters}}, \href{https://doi.org/10.1093/mnras/stx544}{\emph{MNRAS}
  {\bfseries 468} (2017) 1038}
  [\href{https://arxiv.org/abs/1702.08856}{{\ttfamily 1702.08856}}].

\bibitem{Lund}
C.~L. {Sahlholdt}, S.~{Feltzing}, L.~{Lindegren} and R.~P. {Church},
  \emph{{Benchmark ages for the Gaia benchmark stars}},
  \href{https://doi.org/10.1093/mnras/sty2732}{\emph{MNRAS} {\bfseries 482}
  (2019) 895} [\href{https://arxiv.org/abs/1810.02829}{{\ttfamily
  1810.02829}}].

\bibitem{Sarajedini2007}
A.~{Sarajedini}, L.~R. {Bedin}, B.~{Chaboyer}, A.~{Dotter}, M.~{Siegel},
  J.~{Anderson} et~al., \emph{{The ACS Survey of Galactic Globular Clusters. I.
  Overview and Clusters without Previous Hubble Space Telescope Photometry}},
  \href{https://doi.org/10.1086/511979}{\emph{AJ} {\bfseries 133} (2007) 1658}
  [\href{https://arxiv.org/abs/astro-ph/0612598}{{\ttfamily
  astro-ph/0612598}}].

\bibitem{Dotter2010}
A.~{Dotter}, A.~{Sarajedini}, J.~{Anderson}, A.~{Aparicio}, L.~R. {Bedin},
  B.~{Chaboyer} et~al., \emph{{The ACS Survey of Galactic Globular Clusters.
  IX. Horizontal Branch Morphology and the Second Parameter Phenomenon}},
  \href{https://doi.org/10.1088/0004-637X/708/1/698}{\emph{ApJ} {\bfseries 708}
  (2010) 698} [\href{https://arxiv.org/abs/0911.2469}{{\ttfamily 0911.2469}}].

\bibitem{isochrones}
T.~D. {Morton}, \emph{{isochrones: Stellar model grid package}},  Mar., 2015.

\bibitem{MIST0}
A.~{Dotter}, \emph{{MESA Isochrones and Stellar Tracks (MIST) 0: Methods for
  the Construction of Stellar Isochrones}},
  \href{https://doi.org/10.3847/0067-0049/222/1/8}{\emph{ApJS} {\bfseries 222}
  (2016) 8} [\href{https://arxiv.org/abs/1601.05144}{{\ttfamily 1601.05144}}].

\bibitem{MIST1}
J.~{Choi}, A.~{Dotter}, C.~{Conroy}, M.~{Cantiello}, B.~{Paxton} and B.~D.
  {Johnson}, \emph{{Mesa Isochrones and Stellar Tracks (MIST). I. Solar-scaled
  Models}}, \href{https://doi.org/10.3847/0004-637X/823/2/102}{\emph{ApJ}
  {\bfseries 823} (2016) 102}
  [\href{https://arxiv.org/abs/1604.08592}{{\ttfamily 1604.08592}}].

\bibitem{dsed}
A.~{Dotter}, B.~{Chaboyer}, D.~{Jevremovi{\'c}}, V.~{Kostov}, E.~{Baron} and
  J.~W. {Ferguson}, \emph{{The Dartmouth Stellar Evolution Database}},
  \href{https://doi.org/10.1086/589654}{\emph{ApJS} {\bfseries 178} (2008) 89}
  [\href{https://arxiv.org/abs/0804.4473}{{\ttfamily 0804.4473}}].

\bibitem{Simpson70}
E.~{Simpson}, R.~E. {Hills}, W.~{Hoffman}, S.~A. {Kellman}, J.~{Morton}, Erwin,
  F.~{Paresce} et~al., \emph{{Studies in Stellar Evolution.IX. Theoretical
  Isochrones for Early-Type Clusters}},
  \href{https://doi.org/10.1086/150366}{\emph{ApJ} {\bfseries 159} (1970) 895}.

\bibitem{Bertelli90}
G.~{Bertelli}, R.~{Betto}, A.~{Bressan}, C.~{Chiosi}, E.~{Nasi} and
  A.~{Vallenari}, \emph{{Theoretical isochrones with convective overshoot.}},
  {\emph{A\&ASS} {\bfseries 85} (1990) 845}.

\bibitem{Bertelli94}
G.~{Bertelli}, A.~{Bressan}, C.~{Chiosi}, F.~{Fagotto} and E.~{Nasi},
  \emph{{Theoretical isochrones from models with new radiative opacities.}},
  {\emph{A\&ASS} {\bfseries 106} (1994) 275}.

%
  
 \bibitem{Fitz99}
E. L. ~{Fitzpatrick}, \emph{{Correcting for the Effects of Interstellar Extinction}},
 \href{https://iopscience.iop.org/article/10.1086/316293/pdf}{\emph{PASP} {\bfseries 111} (1999) 755}
  [\href{https://arxiv.org/pdf/astro-ph/9809387.pdf}{{\ttfamily astro-ph/9809387}}]. 


\bibitem{ReviewBastian18}
N.~{Bastian} and C.~{Lardo}, \emph{{Multiple Stellar Populations in Globular
  Clusters}},
  \href{https://doi.org/10.1146/annurev-astro-081817-051839}{\emph{ARAA}
  {\bfseries 56} (2018) 83} [\href{https://arxiv.org/abs/1712.01286}{{\ttfamily
  1712.01286}}].

\bibitem{Harris}
W.~E. {Harris}, \emph{{A Catalog of Parameters for Globular Clusters in the
  Milky Way}}, \href{https://doi.org/10.1086/118116}{\emph{Aj} {\bfseries 112}
  (1996) 1487}.

\bibitem{Dutra}
C.~M. {Dutra} and E.~{Bica}, \emph{{Foreground and background dust in star
  cluster directions}}, {\emph{aap} {\bfseries 359} (2000) 347}
  [\href{https://arxiv.org/abs/astro-ph/0005108}{{\ttfamily
  astro-ph/0005108}}].

\bibitem{ngc6121}
B.~{Hendricks}, P.~B. {Stetson}, D.~A. {VandenBerg} and M.~{Dall'Ora}, \emph{{A
  New Reddening Law for M4}},
  \href{https://doi.org/10.1088/0004-6256/144/1/25}{\emph{Aj} {\bfseries 144}
  (2012) 25} [\href{https://arxiv.org/abs/1204.5719}{{\ttfamily 1204.5719}}].

\bibitem{ngc6144}
R.~K. {Neely}, A.~{Sarajedini} and D.~H. {Martins}, \emph{{CCD Photometry of
  the Galactic Globular Cluster NGC 6144}},
  \href{https://doi.org/10.1086/301311}{\emph{Aj} {\bfseries 119} (2000) 1793}.

\bibitem{ngc6723}
J.-W. {Lee}, M.~{L{\'o}pez-Morales}, K.~{Hong}, Y.-W. {Kang}, B.~L. {Pohl} and
  A.~{Walker}, \emph{{Toward a Better Understanding of the Distance Scale from
  RR Lyrae Variable Stars: A Case Study for the Inner Halo Globular Cluster NGC
  6723}}, \href{https://doi.org/10.1088/0067-0049/210/1/6}{\emph{Apjs}
  {\bfseries 210} (2014) 6} [\href{https://arxiv.org/abs/1311.2054}{{\ttfamily
  1311.2054}}].

\bibitem{Carretta}
E.~{Carretta}, A.~{Bragaglia}, R.~{Gratton}, V.~{D'Orazi} and S.~{Lucatello},
  \emph{{Intrinsic iron spread and a new metallicity scale for globular
  clusters}}, \href{https://doi.org/10.1051/0004-6361/200913003}{\emph{aap}
  {\bfseries 508} (2009) 695}
  [\href{https://arxiv.org/abs/0910.0675}{{\ttfamily 0910.0675}}].

\bibitem{Cardelli}
J.~A. {Cardelli}, G.~C. {Clayton} and J.~S. {Mathis}, \emph{{The Relationship
  between Infrared, Optical, and Ultraviolet Extinction}},
  \href{https://doi.org/10.1086/167900}{\emph{ApJ} {\bfseries 345} (1989) 245}.

\bibitem{KosowskyJim}
A.~{Kosowsky}, M.~{Milosavljevic} and R.~{Jimenez}, \emph{{Efficient
  cosmological parameter estimation from microwave background anisotropies}},
  \href{https://doi.org/10.1103/PhysRevD.66.063007}{\emph{PRD} {\bfseries 66}
  (2002) 063007} [\href{https://arxiv.org/abs/astro-ph/0206014}{{\ttfamily
  astro-ph/0206014}}].

\bibitem{Jimenez95}
R.~{Jimenez} and J.~{MacDonald}, \emph{{Stellar evolutionary tracks for
  low-mass stars}}, \href{https://doi.org/10.1093/mnras/283.2.721}{\emph{MNRAS}
  {\bfseries 283} (1996) 721}.

\bibitem{JimGC}
R.~{Jimenez}, A.~{Cimatti}, L.~{Verde}, M.~{Moresco} and B.~{Wandelt},
  \emph{{The local and distant Universe: stellar ages and H$_{0}$}},
  \href{https://doi.org/10.1088/1475-7516/2019/03/043}{\emph{JCAP} {\bfseries
  2019} (2019) 043} [\href{https://arxiv.org/abs/1902.07081}{{\ttfamily
  1902.07081}}].

\bibitem{2018Natur.557..392H}
T.~{Hashimoto}, N.~{Laporte}, K.~{Mawatari}, R.~S. {Ellis}, A.~K. {Inoue},
  E.~{Zackrisson} et~al., \emph{{The onset of star formation 250 million years
  after the Big Bang}},
  \href{https://doi.org/10.1038/s41586-018-0117-z}{\emph{Nature} {\bfseries
  557} (2018) 392} [\href{https://arxiv.org/abs/1805.05966}{{\ttfamily
  1805.05966}}].

\bibitem{2020ApJ...888..124S}
V.~{Strait}, M.~{Brada{\v{c}}}, D.~{Coe}, L.~{Bradley}, B.~{Salmon}, B.~C.
  {Lemaux} et~al., \emph{{Stellar Properties of z {\ensuremath{\gtrsim}} 8
  Galaxies in the Reionization Lensing Cluster Survey}},
  \href{https://doi.org/10.3847/1538-4357/ab5daf}{\emph{ApJ} {\bfseries 888}
  (2020) 124} [\href{https://arxiv.org/abs/1905.09295}{{\ttfamily
  1905.09295}}].

\bibitem{2019MNRAS.489.3827B}
C.~{Binggeli}, E.~{Zackrisson}, X.~{Ma}, A.~K. {Inoue}, A.~{Vikaeus},
  T.~{Hashimoto} et~al., \emph{{Balmer breaks in simulated galaxies at $z >
  6$}}, \href{https://doi.org/10.1093/mnras/stz2387}{\emph{MNRAS} {\bfseries
  489} (2019) 3827} [\href{https://arxiv.org/abs/1908.11393}{{\ttfamily
  1908.11393}}].
  
\bibitem{Padoan_Jimenez1997}
P.~{Padoan}, R.~{Jimenez}, Raul and B.~{Jones}, \emph{{On star formation in primordial protoglobular clouds}}, \href{https://doi.org/10.1093/mnras/285.4.711}{\emph{MNRAS} {\bfseries
  285} (1997) 711} [\href{https://arxiv.org/abs/astro-ph/9604055}{{\ttfamily
  astro-ph/9604055}}].
  
\bibitem{Trenti2015}
M.~{Trenti}, P.~{Padoan}, R.~{Jimenez}, \emph{{The Relative and Absolute Ages of Old Globular Clusters in the LCDM Framework}}, \href{https://10.1088/2041-8205/808/2/L35}{\emph{ApJL}
  {\bfseries 808} (2015) L35}
  [\href{https://arxiv.org/abs/1502.02670}{{\ttfamily 1502.02670}}].


\bibitem{Choksi2018}
N.~ {Choksi}, O.~{Gnedin} and H.~{Li},
   \emph{Formation of globular cluster systems: from dwarf galaxies to giants},\href{http://dx.doi.org/10.1093/mnras/sty1952}{\emph{MNRAS} {\bfseries
  480} (2018) 2343} [\href{https://arxiv.org/abs/1801.03515}{{\ttfamily
  1801.03515}}].
  
\bibitem{Reina_Campos2019}
M.~{Reina-Campos},\emph{{The origin of metal-poor and metal-rich globular clusters in E-MOSAICS}}, \href{http://dx.doi.org/10.1017/S1743921319006884}, \emph{{Proceedings of the International Astronomical Union}{\bfseries 14} (2019) 147}  [\href{https://arxiv.org/abs/astro-ph/0109232}{{\ttfamily  astro-ph/0109232}}].

\bibitem{Kruijssen2019}
J. M. D.~{Kruijssen}, \emph{{The minimum metallicity of globular clusters and its physical origin – implications for the galaxy mass–metallicity relation and observations of proto-globular clusters at high redshift}},  \href{http://dx.doi.org/10.1093/mnrasl/slz052}{\emph{MNRASL}{\bfseries 486} (2019) L20}  [\href{https://arxiv.org/abs/1904.09987}{{\ttfamily  1904.09987}}].

\bibitem{Forbes2015}
D. A.~{Forbes}, N.~{Pastorello}, A.J.~{Romanowsky}, C.~{Usher}, J.P.~{Brodie}, J.~{Strader}, \emph{{SLUGGS survey: inferring the formation epochs of metal-poor and metal-rich globular clusters}},\href{http://dx.doi.org/10.1093/mnras/stv1312}{\emph{MNRASL}{\bfseries 452} (2015) 1045}  [\href{https://arxiv.org/abs/1506.06820}{{\ttfamily  1506.06820}}].

  
\bibitem{Planck18}
{Planck Collaboration}, N.~{Aghanim}, Y.~{Akrami}, M.~{Ashdown}, J.~{Aumont},
  C.~{Baccigalupi} et~al., \emph{{Planck 2018 results. VI. Cosmological
  parameters}}, {\emph{arXiv e-prints} (2018) arXiv:1807.06209}
  [\href{https://arxiv.org/abs/1807.06209}{{\ttfamily 1807.06209}}].

\bibitem{Knox}
L.~{Knox}, N.~{Christensen} and C.~{Skordis}, \emph{{The Age of the Universe
  and the Cosmological Constant Determined from Cosmic Microwave Background
  Anisotropy Measurements}}, \href{https://doi.org/10.1086/338655}{\emph{ApJL}
  {\bfseries 563} (2001) L95}
  [\href{https://arxiv.org/abs/astro-ph/0109232}{{\ttfamily
  astro-ph/0109232}}].

\bibitem{KW}
R.~{Kippenhahn} and A.~{Weigert}, \emph{{Stellar Structure and Evolution}}.
  1990.

\end{thebibliography}

\providecommand{\href}[2]{#2}\begingroup\raggedright\endgroup

\clearpage

\appendix

\section{Test of sensitivity of the  color-magnitude diagram to model parameters}
\label{sec:sensitivity}

In this appendix we explore the dependence of the isochrones in the color-magnitude diagram of a GC on the model parameters. On top of illustrating the sensitivity of different sections of the evolutionary track to these parameters, this exercise will allow us  to convey how parameter degeneracies can be lifted by considering  regions above the main sequence.  We start from a common set of parameters (based on estimates from literature, see Ref.~\citep{Dotter2010}) and vary one parameter  at a time, while we keep the others fixed. As we vary the parameter of interest, we compare the color at the interpolated magnitudes for each isochrone.
 
\begin{figure}[t!]
\centering
\includegraphics[scale=0.41]{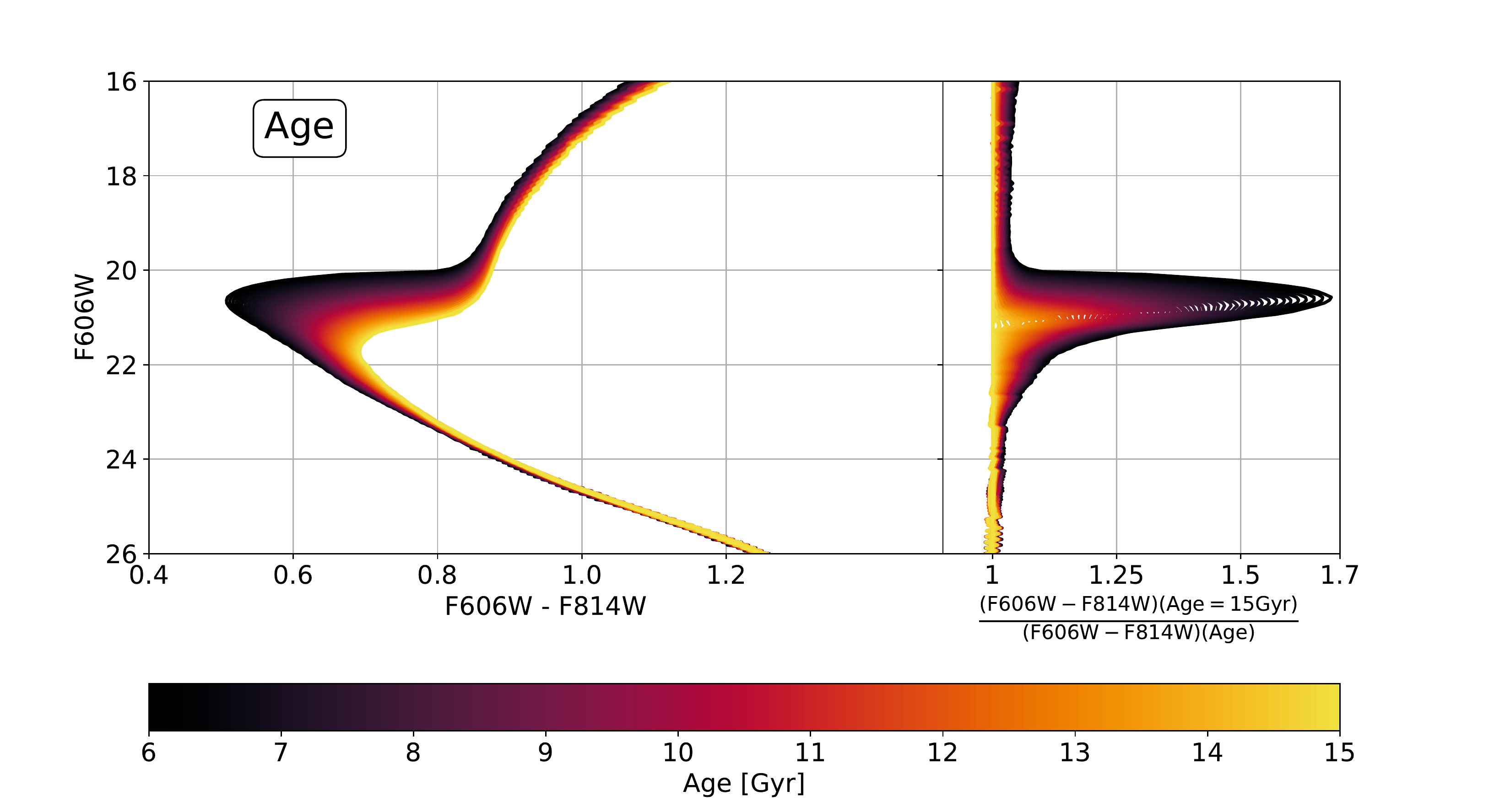} 
\includegraphics[scale=0.41]{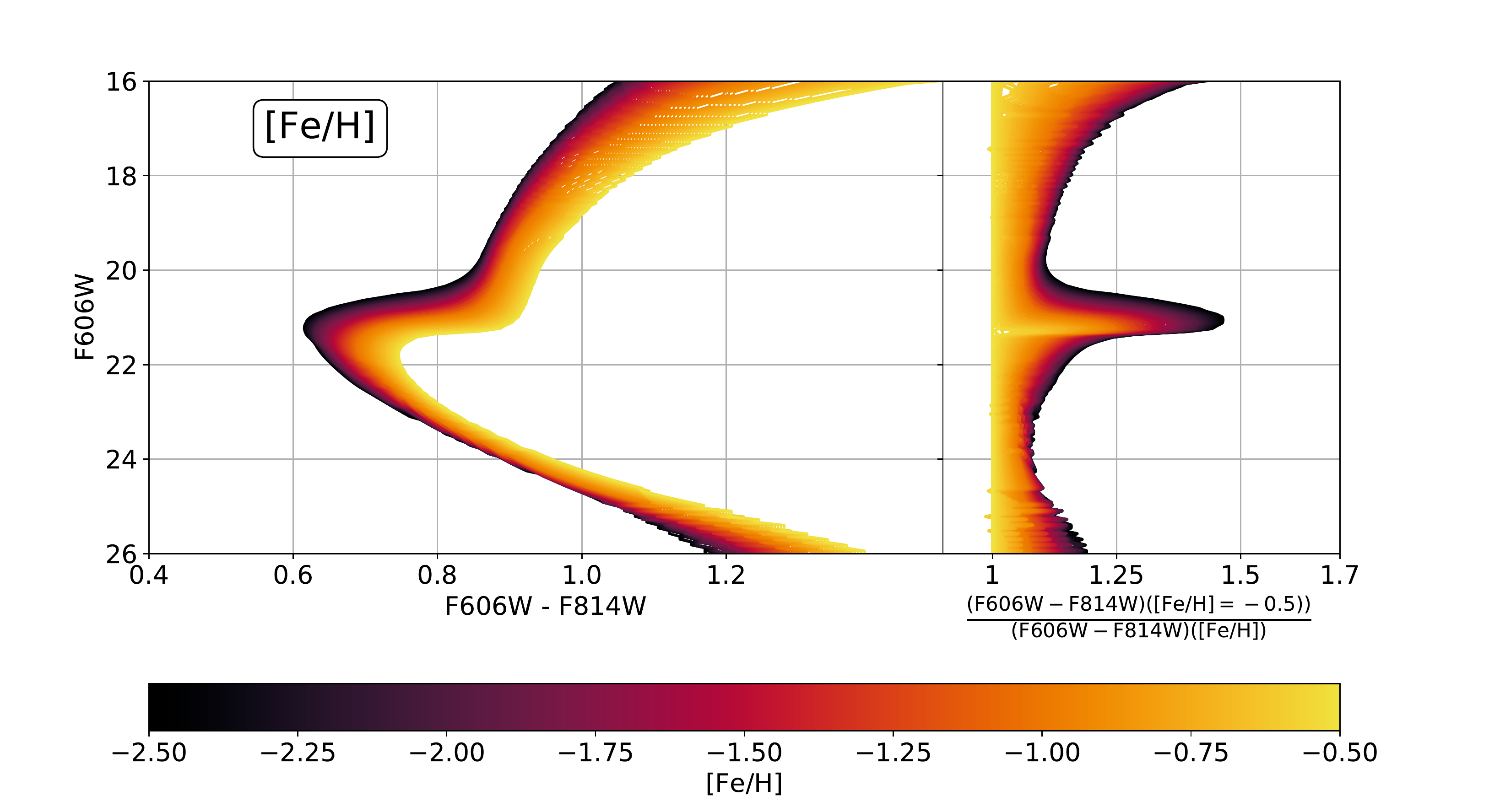}
\includegraphics[scale=0.41]{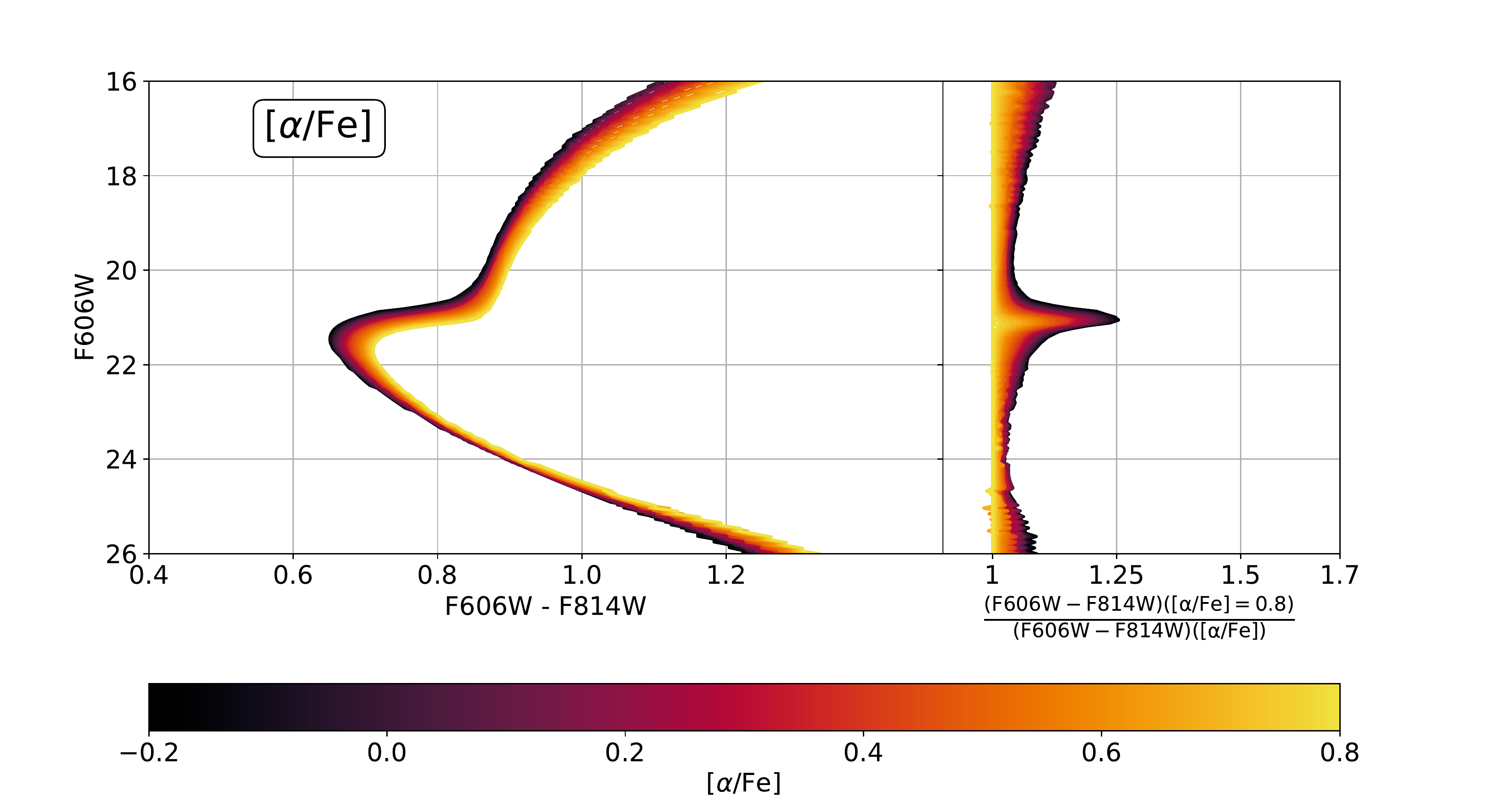} 
\caption{Dependence of the stellar isochrone on variations of age, metallicity and [$\alpha$/Fe] of the GC with all other parameters fixed. Right panels show the relative difference in color.}
\label{diff1}
\end{figure}

  We show the corresponding comparison as function of age, metallicity, and [$\alpha$/Fe] in Figure~\ref{diff1}. The figure  clearly shows that most of the  sensitivity to age is around the MSTOP, but if only this point is used,  age is  degenerate with metallicity. However, both the red giant branch and the lower main sequence are sensitive to metallicity, significantly more than to age. This explains why using more features of the color-magnitude diagram breaks the degeneracy. Further, the whole color-magnitude diagram has a different sensitivity to [$\alpha$/Fe] than to [Fe/H]. Thus, with enough signal-to-noise,  both quantities can be constrained in a joint analysis.

\clearpage

\clearpage
\section{Globular clusters properties after the cuts}

\label{app:GCtable}
\begin{table}[h!]
\centering
\resizebox{12cm}{10cm}{%
\begin{tabular}{|c|c|c|c|}
\hline
Cluster name & Total number of stars & Stars with magnitude  $ < m_{cut}$ & percentage of remaining stars\\ \hline
arp2 & 23010 & 10611 & 46\\ \hline 
ic4499 & 61931 & 33938 & 54 \\ \hline
lynga7 & 44927 & 27496 & 61 \\ \hline
ngc0104 & 140016 & 113700 & 81 \\ \hline
ngc0288 & 26814 & 14465 & 53 \\ \hline
ngc0362 & 111393 & 71978 & 64 \\ \hline
ngc1261 & 97780 & 61767 & 63 \\ \hline
ngc1851 & 130655 & 82732 & 63 \\ \hline
ngc2298 & 20288 & 13453 & 66 \\ \hline
ngc2808 & 277727 & 214443 & 77 \\ \hline
ngc3201 & 31908 & 17056 & 53 \\ \hline
ngc4147 & 19717 & 13977 & 70 \\ \hline
ngc4590 & 60058 & 33182 & 55 \\ \hline
ngc4833 & 60889 & 41720 & 68 \\ \hline
ngc5024 & 222899 & 132605 & 59 \\ \hline
ngc5053 & 23957 & 11104 & 46 \\ \hline
ngc5139 & 300622 & 206535 & 68 \\ \hline
ngc5272 & 161342 & 106494 & 66 \\ \hline
ngc5286 & 190379 & 131490 & 69 \\ \hline
ngc5466 & 29776 & 13660 & 45 \\ \hline
ngc5904 & 108602 & 73235 & 67 \\ \hline
ngc5927 & 96349 & 69333 & 71 \\ \hline
ngc5986 & 148963 & 100314 & 67 \\ \hline
ngc6093 & 125128 & 88784 & 70 \\ \hline
ngc6101 & 67032 & 33715 & 50 \\ \hline
ngc6121 & 11975 & 7070 & 59 \\ \hline
ngc6144 & 22485 & 15612 & 69 \\ \hline
ngc6205 & 138295 & 97673 & 70 \\ \hline
ngc6218 & 29767 & 20840 & 70 \\ \hline
ngc6254 & 54662 & 38462 & 70 \\ \hline
ngc6304 & 100830 & 58706 & 58 \\ \hline
ngc6341 & 129969 & 83376 & 64 \\ \hline
ngc6352 & 25779 & 14784 & 57 \\ \hline
ngc6362 & 30541 & 17724 & 58 \\ \hline
ngc6366 & 10567 & 4427 & 41 \\ \hline
ngc6388 & 310630 & 257049 & 82 \\ \hline
ngc6397 & 14277 & 9404 & 65 \\ \hline
ngc6426 & 57321 & 30576 & 53 \\ \hline
ngc6441 & 340872 & 299187 & 87 \\ \hline
ngc6496 & 22938 & 14486 & 63 \\ \hline
ngc6535 & 9590 & 3640 & 37 \\ \hline
ngc6541 & 111010 & 71816 & 64 \\ \hline
ngc6584 & 62694 & 35346 & 56 \\ \hline
ngc6624 & 62637 & 40960 & 65 \\ \hline
ngc6637 & 61801 & 44484 & 71 \\ \hline
ngc6652 & 29936 & 16586 & 55 \\ \hline
ngc6656 & 92090 & 57379 & 62 \\ \hline
ngc6681 & 48442 & 32417 & 66 \\ \hline
ngc6715 & 345989 & 270157 & 78 \\ \hline
ngc6717 & 15209 & 8235 & 54 \\ \hline
ngc6723 & 60289 & 42353 & 70 \\ \hline
ngc6752 & 47657 & 31250 & 65 \\ \hline
ngc6779 & 79381 & 47224 & 59 \\ \hline
ngc6809 & 42870 & 24095 & 56 \\ \hline
ngc6838 & 14504 & 7582 & 52 \\ \hline
ngc6934 & 81104 & 47218 & 58 \\ \hline
ngc6981 & 44154 & 29154 & 66 \\ \hline
ngc7006 & 72056 & 46216 & 64 \\ \hline
ngc7078 & 243929 & 152629 & 62 \\ \hline
ngc7089 & 227533 & 159739 & 70 \\ \hline
ngc7099 & 67053 & 37756 & 56 \\ \hline
palomar1 & 9330 & 685 & 7 \\ \hline
palomar12 & 7915 & 1981 & 25 \\ \hline
palomar15 & 22790 & 6648 & 29 \\ \hline
pyxis & 11311 & 6281 & 55 \\ \hline
ruprecht106 & 23800 & 13285 & 55 \\ \hline
terzan7 & 21637 & 7752 & 35 \\ \hline
terzan8 & 39847 & 16477 & 41 \\ \hline
\end{tabular}%
}
\caption{Impact of the magnitude cut on the number of stars; All numbers are given after the photometry cleaning}
\label{tab:stars_number}
\end{table}

\clearpage
\section{Main sequence calibration}
\label{app:MScalib}
\begin{figure}[h!]
\centering
\includegraphics[scale=0.3]{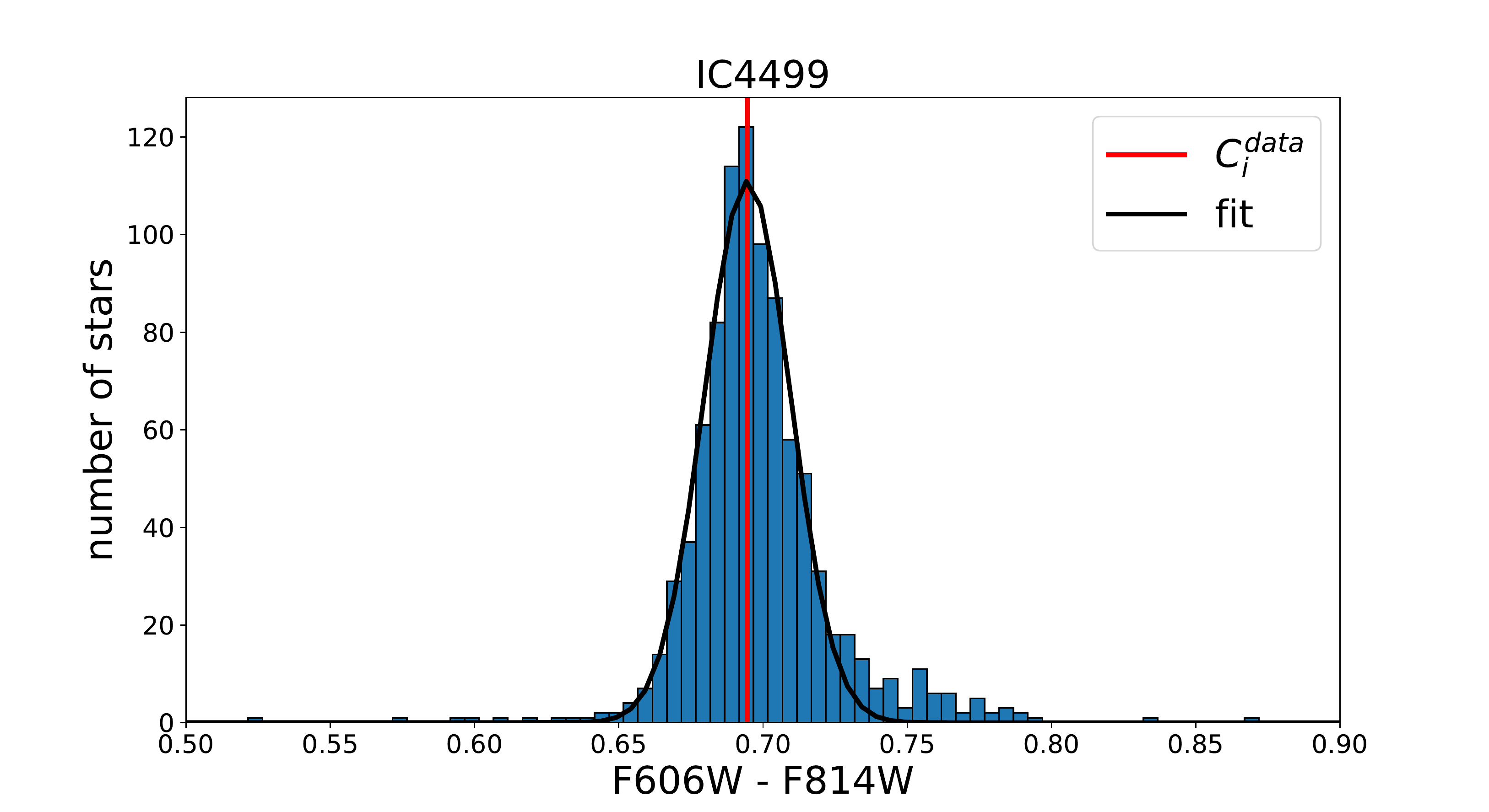} 
\caption{Distribution of color inside a typical MS magnitude bin (at the dim end of the MS) showing  secondary population contamination. The black solid line shows how the algorithm isolates and fits the distribution of the  main population. }
\label{fitms2}
\end{figure}

We fit the histogram of the color distribution within each magnitude bin with a unimodal Gaussian with the \texttt{curve\_fit} routine of \texttt{Scipy}, for a \texttt{maxfev}=950000.\footnote{\texttt{maxfev} is set to a very large value to make sure of the non convergence of the unimodal fit.} If the routine cannot find a fit to the color distribution,  the bin is ignored. Otherwise,  the bin is retained and the  resulting Gaussian distribution is adopted. A typical example  of a small contamination  is shown in Figure ~\ref{fitms2}: the fitting procedure  captures the distribution of the  main ``population". 

\begin{figure}[h!]
\centering
\includegraphics[scale=0.3]{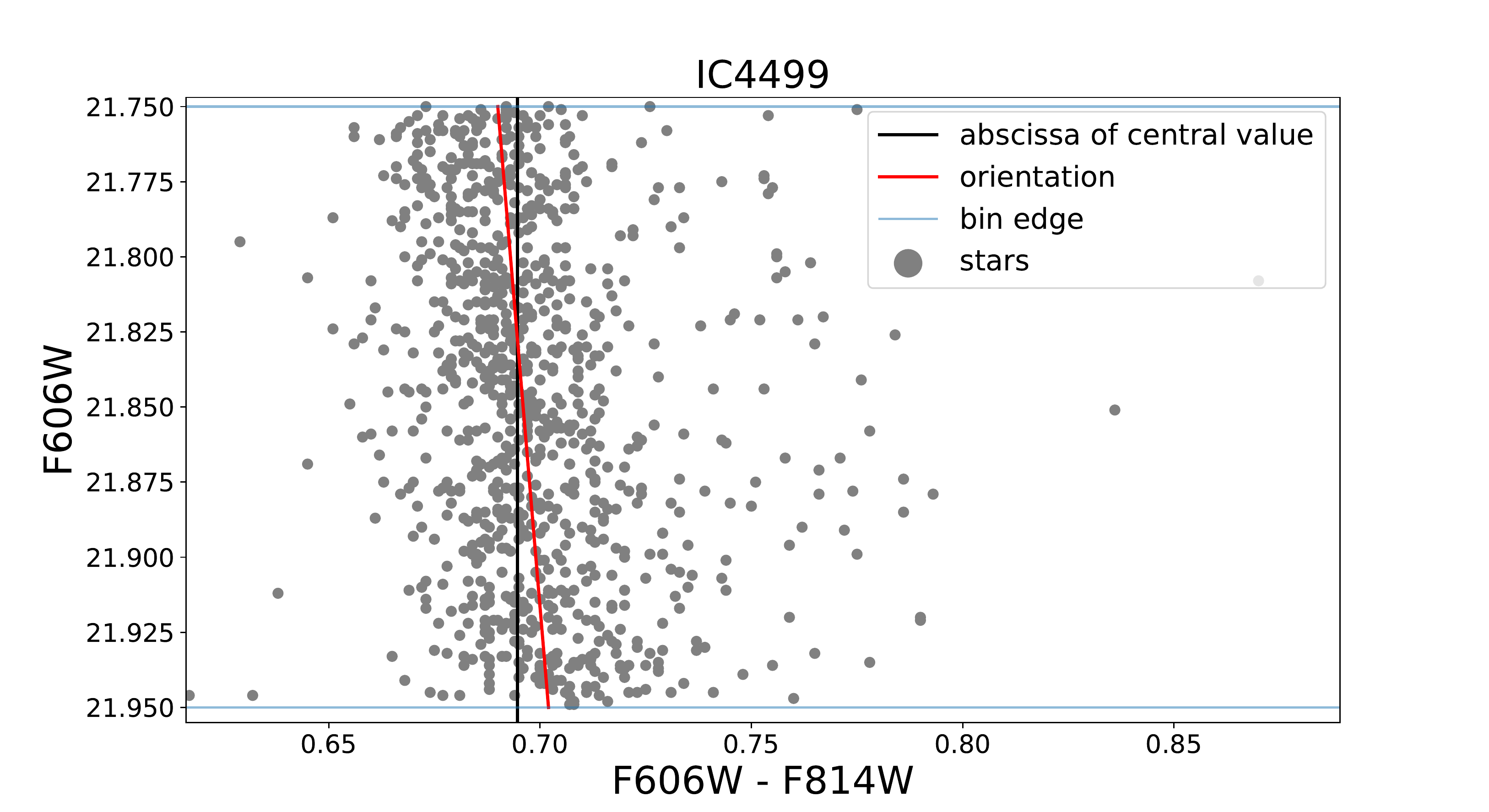} 
\caption{Orientation of the data compared to the vertical axis inside a typical MS magnitude bin (far away from the dim end cut).}
\label{fitms1}
\end{figure}

Once the central value of the distribution is obtained (see Figure ~\ref{fitms2}), we rescale the error on the distribution due to the inclination of the observed stars in the color-magnitude diagram. The orientation of the data in a MS magnitude bin $i$  is obtained by linear regression the median of the data in sub-bins, and is compared to a vertical line passing through the color of the central value (see Figure ~\ref{fitms1}). The resulting angle is referred to as $\phi_i$ and it ranges between [ $\simeq$ 0 - 10$^\circ$]. 

\newpage
\section{Mixing Length Theory}
\label{appendix:MLT}

Uncertainties in the modeling of convection in the envelopes of low mass stars are the main contributor to systematic uncertainties in determining stellar parameters (see Table~2 in Ref.~\cite{OMalley}).  Given the broad audience which this paper (hopefully) reaches,  it is worth to briefly review mixing length theory (MLT), to understand the origin of these uncertainties.

The envelopes (about the outer 30\% radius) of low mass ($< 2$ M$_{\odot}$) stars are fully convective and turbulent, with Reynolds number $\approx 10^{10}$. Modeling these systems is highly challenging:
 in principle, a full hydro-dynamical solution should be obtained.
Instead, the standard solution is to model the gradient of convective transport by the so-called MLT. Conceptually, it is a very simple approach: it assumes that a blob of gas starts at a point and continues moving until it dissolves after a certain length, the mixing length $l_m$.

Consider a sphere of radius $r$ and an element $e$ of the envelope (a blob of gas) located there. After $e$  has traveled a mixing length $l_m$, its increase in temperature $T$ will be
\begin{equation}
\frac{\Delta T}{T} = \frac{1}{T} \frac{\partial (\Delta T)}{\partial r} l_m = (\nabla - \nabla_e) l_m \frac{1}{H_P}
\end{equation}
where the scale height is $H_P = -dr/d \ln P$, $P$ is the pressure, $\nabla$ denotes the gradient in the environment  and $\nabla_e$ is the gradient in the blob. Now, combining this with the equations of stellar structure, it is possible to obtain a system of five differential equations for five independent variables, namely: pressure, temperature, density, and the advective and radiative gradients. Then  $l_m$  is  an extra free parameter which needs to be determined from observations. The usual parameter that stellar modelers fit is $\alpha_{\rm MLT} = l_m/H_P$; this has a typical value of $1.6$ from fits to the Sun and   to the position of red giant branch in the color-magnitude diagram of GCs~\cite{JimenezGC96}. The interested reader can consult the textbook by Kippenhahn \& Weigert for a detailed account of all equations of stellar structure~\cite{KW}. Changes for this parameter from the typical value would propagate into systematic shifts in the  metallicity and age determinations.

\newpage
\section{Parameter constraints: globular clusters}
\label{app:GCtable-params}

The table shows the best-fit parameters for the GC sample considered in this paper and the  one-dimensional marginalized statistical uncertainties at 68\% confidence level.

\begin{table}[h!]
\scriptsize
\centering
\renewcommand{\arraystretch}{1.6}
\begin{tabular}{|c|c|c|c|c|c|}
\hline
Cluster name & Age [Gyr] & [Fe/H] & Distance [kpc] & $A_{V}$ & [$\alpha$/Fe] \\ \hline
arp2 & $13.42^{+1.24}_{-1.65}$ & $-1.81^{+0.15}_{-0.18}$ & $29.61^{+1.46}_{-1.46}$ & $0.29^{+0.04}_{-0.04}$ & $0.14^{+0.18}_{-0.16}$\\ \hline
ic4499 & $12.80^{+0.66}_{-0.78}$ & $-1.54^{+0.09}_{-0.14}$ & $19.68^{+0.45}_{-0.45}$ & $0.64^{+0.03}_{-0.03}$ & $-0.09^{+0.17}_{-0.10}$\\ \hline
lynga7 & $10.82^{+2.12}_{-1.54}$ & $-0.88^{+0.15}_{-0.12}$ & $9.23^{+0.58}_{-0.58}$ & $2.22^{+0.05}_{-0.05}$ & $0.03^{+0.18}_{-0.12}$\\ \hline
ngc0104 & $13.54^{+1.03}_{-0.80}$ & $-0.81^{+0.10}_{-0.14}$ & $4.48^{+0.08}_{-0.11}$ & $0.06^{+0.03}_{-0.03}$ & $0.20^{+0.12}_{-0.16}$\\ \hline
ngc0288 & $11.20^{+0.67}_{-0.67}$ & $-1.43^{+0.18}_{-0.11}$ & $9.77^{+0.20}_{-0.20}$ & $0.05^{+0.04}_{-0.03}$ & $0.36^{+0.16}_{-0.18}$\\ \hline
ngc0362 & $11.52^{+0.84}_{-0.84}$ & $-1.30^{+0.14}_{-0.12}$ & $9.12^{+0.21}_{-0.21}$ & $0.06^{+0.03}_{-0.03}$ & $0.12^{+0.16}_{-0.14}$\\ \hline
ngc1261 & $11.54^{+0.67}_{-0.45}$ & $-1.32^{+0.12}_{-0.12}$ & $16.72^{+0.38}_{-0.25}$ & $0.00^{+0.03}_{0.00}$ & $0.12^{+0.16}_{-0.14}$\\ \hline
ngc1851 & $12.27^{+1.47}_{-0.98}$ & $-1.14^{+0.13}_{-0.13}$ & $12.19^{+0.32}_{-0.32}$ & $0.07^{+0.03}_{-0.04}$ & $0.05^{+0.14}_{-0.14}$\\ \hline
ngc2298 & $13.89^{+0.88}_{-0.63}$ & $-2.00^{+0.14}_{-0.16}$ & $10.23^{+0.22}_{-0.22}$ & $0.65^{+0.03}_{-0.03}$ & $0.18^{+0.12}_{-0.20}$\\ \hline
ngc2808 & $10.93^{+1.20}_{-1.03}$ & $-1.32^{+0.14}_{-0.12}$ & $10.84^{+0.38}_{-0.38}$ & $0.60^{+0.04}_{-0.04}$ & $-0.06^{+0.09}_{-0.13}$\\ \hline
ngc3201 & $13.05^{+1.05}_{-1.19}$ & $-1.57^{+0.12}_{-0.15}$ & $4.91^{+0.15}_{-0.08}$ & $0.74^{+0.04}_{-0.03}$ & $0.20^{+0.16}_{-0.14}$\\ \hline
ngc4147 & $13.02^{+0.50}_{-0.33}$ & $-1.77^{+0.14}_{-0.12}$ & $19.58^{+0.34}_{-0.34}$ & $0.04^{+0.02}_{-0.02}$ & $0.22^{+0.18}_{-0.16}$\\ \hline
ngc4590 & $12.03^{+0.54}_{-0.54}$ & $-2.28^{+0.17}_{-0.11}$ & $11.22^{+0.17}_{-0.25}$ & $0.18^{+0.02}_{-0.02}$ & $0.26^{+0.16}_{-0.22}$\\ \hline
ngc4833 & $14.69^{+0.23}_{-0.70}$ & $-2.09^{+0.15}_{-0.15}$ & $6.91^{+0.18}_{-0.12}$ & $0.97^{+0.02}_{-0.03}$ & $0.20^{+0.18}_{-0.14}$\\ \hline
ngc5024 & $13.31^{+0.66}_{-0.57}$ & $-1.95^{+0.11}_{-0.17}$ & $18.99^{+0.55}_{-0.37}$ & $0.03^{+0.01}_{-0.02}$ & $0.34^{+0.14}_{-0.18}$\\ \hline
ngc5053 & $13.84^{+0.50}_{-0.58}$ & $-2.33^{+0.14}_{-0.12}$ & $17.82^{+0.29}_{-0.29}$ & $0.02^{+0.01}_{-0.01}$ & $0.16^{+0.16}_{-0.18}$\\ \hline
ngc5139 & $14.91^{+0.00}_{-1.11}$ & $-1.63^{+0.11}_{-0.14}$ & $5.78^{+0.16}_{-0.16}$ & $0.36^{+0.03}_{-0.02}$ & $-0.00^{+0.11}_{-0.13}$\\ \hline
ngc5272 & $12.60^{+0.66}_{-0.66}$ & $-1.53^{+0.11}_{-0.13}$ & $10.41^{+0.18}_{-0.28}$ & $0.00^{+0.02}_{0.00}$ & $0.26^{+0.12}_{-0.22}$\\ \hline
ngc5286 & $14.55^{+0.36}_{-1.07}$ & $-1.71^{+0.15}_{-0.15}$ & $11.62^{+0.40}_{-0.27}$ & $0.73^{+0.03}_{-0.03}$ & $-0.01^{+0.14}_{-0.14}$\\ \hline
ngc5466 & $12.31^{+0.60}_{-0.40}$ & $-1.85^{+0.14}_{-0.12}$ & $16.47^{+0.39}_{-0.17}$ & $0.00^{+0.02}_{-0.00}$ & $0.30^{+0.18}_{-0.18}$\\ \hline
ngc5904 & $12.75^{+0.80}_{-0.80}$ & $-1.30^{+0.10}_{-0.16}$ & $7.53^{+0.11}_{-0.17}$ & $0.07^{+0.02}_{-0.03}$ & $0.12^{+0.14}_{-0.16}$\\ \hline
ngc5927 & $8.33^{+1.98}_{-1.13}$ & $-0.62^{+0.13}_{-0.13}$ & $8.87^{+0.20}_{-0.39}$ & $1.35^{+0.03}_{-0.07}$ & $0.16^{+0.12}_{-0.14}$\\ \hline
ngc5986 & $14.82^{+0.00}_{-1.12}$ & $-1.66^{+0.13}_{-0.16}$ & $10.95^{+0.40}_{0.00}$ & $0.82^{+0.03}_{-0.03}$ & $0.19^{+0.06}_{-0.22}$\\ \hline
ngc6093 & $13.83^{+0.96}_{-0.72}$ & $-1.79^{+0.16}_{-0.13}$ & $10.97^{+0.26}_{-0.26}$ & $0.61^{+0.03}_{-0.03}$ & $0.20^{+0.14}_{-0.18}$\\ \hline
ngc6101 & $13.22^{+0.66}_{-0.66}$ & $-1.84^{+0.15}_{-0.12}$ & $14.81^{+0.37}_{-0.25}$ & $0.30^{+0.03}_{-0.03}$ & $0.30^{+0.16}_{-0.16}$\\ \hline
ngc6121 & $13.01^{+1.01}_{-1.01}$ & $-1.22^{+0.16}_{-0.09}$ & $2.05^{+0.03}_{-0.05}$ & $1.15^{+0.02}_{-0.02}$ & $0.44^{+0.12}_{-0.10}$\\ \hline
ngc6144 & $14.47^{+0.42}_{-1.12}$ & $-1.76^{+0.15}_{-0.17}$ & $8.72^{+0.23}_{-0.23}$ & $1.27^{+0.03}_{-0.03}$ & $0.22^{+0.16}_{-0.18}$\\ \hline
ngc6205 & $13.49^{+0.62}_{-0.45}$ & $-1.48^{+0.08}_{-0.16}$ & $7.79^{+0.09}_{-0.12}$ & $0.00^{+0.02}_{0.00}$ & $0.08^{+0.27}_{-0.08}$\\ \hline
ngc6218 & $14.64^{+0.29}_{-0.64}$ & $-1.51^{+0.13}_{-0.11}$ & $5.27^{+0.12}_{-0.04}$ & $0.56^{+0.03}_{-0.03}$ & $0.28^{+0.10}_{-0.22}$\\ \hline
ngc6254 & $12.85^{+0.80}_{-0.80}$ & $-1.75^{+0.13}_{-0.13}$ & $5.71^{+0.14}_{-0.12}$ & $0.78^{+0.04}_{-0.03}$ & $0.09^{+0.10}_{-0.22}$\\ \hline
ngc6304 & $8.67^{+1.80}_{-1.80}$ & $-0.64^{+0.14}_{-0.12}$ & $7.20^{+0.35}_{-0.35}$ & $1.56^{+0.05}_{-0.05}$ & $0.09^{+0.14}_{-0.14}$\\ \hline
\end{tabular}%
\caption{}
\label{tab:my-table}
\end{table}

\begin{table}[h!]
\scriptsize
\centering
\renewcommand{\arraystretch}{1.6}
\begin{tabular}{|c|c|c|c|c|c|}
\hline
Cluster name & Age [Gyr] & [Fe/H] & Distance [kpc] & $A_{V}$ & [$\alpha$/Fe] \\ \hline
ngc6341 & $13.30^{+0.60}_{-0.60}$ & $-2.24^{+0.15}_{-0.13}$ & $8.94^{+0.20}_{-0.17}$ & $0.05^{+0.03}_{-0.03}$ & $0.26^{+0.16}_{-0.18}$\\ \hline
ngc6352 & $11.93^{+1.80}_{-1.57}$ & $-0.82^{+0.20}_{-0.09}$ & $5.64^{+0.23}_{-0.18}$ & $0.72^{+0.04}_{-0.04}$ & $0.26^{+0.18}_{-0.12}$\\ \hline
ngc6362 & $13.58^{+0.82}_{-0.61}$ & $-1.11^{+0.14}_{-0.11}$ & $7.69^{+0.18}_{-0.08}$ & $0.16^{+0.02}_{-0.02}$ & $0.34^{+0.14}_{-0.08}$\\ \hline
ngc6366 & $12.15^{+1.46}_{-1.46}$ & $-0.88^{+0.15}_{-0.12}$ & $3.68^{+0.11}_{-0.11}$ & $2.13^{+0.05}_{-0.03}$ & $0.03^{+0.16}_{-0.14}$\\ \hline
ngc6388 & $11.07^{+2.12}_{-1.42}$ & $-0.79^{+0.16}_{-0.11}$ & $12.65^{+0.49}_{-0.61}$ & $1.09^{+0.04}_{-0.04}$ & $-0.09^{+0.11}_{-0.10}$\\ \hline
ngc6397 & $14.21^{+0.69}_{-0.69}$ & $-2.06^{+0.14}_{-0.19}$ & $2.65^{+0.05}_{-0.05}$ & $0.51^{+0.03}_{-0.02}$ & $0.18^{+0.18}_{-0.16}$\\ \hline
ngc6426 & $13.92^{+0.96}_{-1.12}$ & $-2.16^{+0.18}_{-0.18}$ & $21.99^{+0.85}_{-1.02}$ & $1.12^{+0.04}_{-0.04}$ & $-0.05^{+0.20}_{-0.12}$\\ \hline
ngc6441 & $10.44^{+2.78}_{-1.62}$ & $-0.65^{+0.12}_{-0.12}$ & $14.38^{+1.15}_{-0.92}$ & $1.42^{+0.06}_{-0.05}$ & $-0.11^{+0.12}_{-0.09}$\\ \hline
ngc6496 & $10.86^{+2.11}_{-1.64}$ & $-0.57^{+0.08}_{-0.14}$ & $9.88^{+0.44}_{-0.29}$ & $0.68^{+0.05}_{-0.04}$ & $0.12^{+0.12}_{-0.12}$\\ \hline
ngc6535 & $13.81^{+1.06}_{-1.06}$ & $-1.82^{+0.18}_{-0.18}$ & $6.71^{+0.38}_{-0.31}$ & $1.22^{+0.02}_{-0.03}$ & $0.20^{+0.14}_{-0.20}$\\ \hline
ngc6541 & $13.51^{+0.86}_{-0.65}$ & $-1.98^{+0.16}_{-0.12}$ & $7.97^{+0.18}_{-0.18}$ & $0.35^{+0.04}_{-0.03}$ & $0.30^{+0.18}_{-0.16}$\\ \hline
ngc6584 & $12.72^{+0.76}_{-0.66}$ & $-1.56^{+0.12}_{-0.14}$ & $14.18^{+0.24}_{-0.37}$ & $0.25^{+0.03}_{-0.03}$ & $0.18^{+0.14}_{-0.18}$\\ \hline
ngc6624 & $11.26^{+1.90}_{-1.27}$ & $-0.61^{+0.10}_{-0.12}$ & $8.53^{+0.33}_{-0.41}$ & $0.80^{+0.05}_{-0.03}$ & $-0.10^{+0.11}_{-0.09}$\\ \hline
ngc6637 & $12.85^{+1.35}_{-1.35}$ & $-0.84^{+0.15}_{-0.10}$ & $9.09^{+0.29}_{-0.29}$ & $0.50^{+0.04}_{-0.04}$ & $0.08^{+0.14}_{-0.16}$\\ \hline
ngc6652 & $12.98^{+1.55}_{-0.86}$ & $-0.86^{+0.12}_{-0.14}$ & $9.57^{+0.39}_{-0.39}$ & $0.35^{+0.03}_{-0.03}$ & $0.18^{+0.14}_{-0.20}$\\ \hline
ngc6656 & $14.54^{+0.36}_{-0.97}$ & $-1.70^{+0.15}_{-0.15}$ & $3.62^{+0.09}_{-0.09}$ & $1.04^{+0.03}_{-0.03}$ & $0.03^{+0.12}_{-0.16}$\\ \hline
ngc6681 & $13.87^{+0.73}_{-0.83}$ & $-1.68^{+0.14}_{-0.14}$ & $9.66^{+0.22}_{-0.27}$ & $0.29^{+0.03}_{-0.03}$ & $0.16^{+0.16}_{-0.16}$\\ \hline
ngc6715 & $12.22^{+1.91}_{-1.43}$ & $-1.54^{+0.13}_{-0.19}$ & $28.25^{+1.58}_{-1.58}$ & $0.44^{+0.04}_{-0.04}$ & $-0.04^{+0.14}_{-0.14}$\\ \hline
ngc6717 & $11.65^{+1.50}_{-1.71}$ & $-1.29^{+0.15}_{-0.15}$ & $7.91^{+0.60}_{-0.34}$ & $0.66^{+0.05}_{-0.05}$ & $0.20^{+0.12}_{-0.20}$\\ \hline
ngc6723 & $13.81^{+0.70}_{-0.90}$ & $-1.06^{+0.07}_{-0.15}$ & $8.14^{+0.16}_{-0.16}$ & $0.20^{+0.03}_{-0.03}$ & $0.24^{+0.09}_{-0.13}$\\ \hline
ngc6752 & $13.48^{+0.81}_{-0.54}$ & $-1.57^{+0.14}_{-0.14}$ & $4.34^{+0.09}_{-0.06}$ & $0.13^{+0.03}_{-0.03}$ & $0.26^{+0.14}_{-0.18}$\\ \hline
ngc6779 & $14.85^{+0.08}_{-0.76}$ & $-2.13^{+0.14}_{-0.16}$ & $10.92^{+0.27}_{-0.18}$ & $0.69^{+0.02}_{-0.02}$ & $0.09^{+0.14}_{-0.16}$\\ \hline
ngc6809 & $13.93^{+0.50}_{-0.58}$ & $-1.80^{+0.11}_{-0.11}$ & $5.49^{+0.09}_{-0.07}$ & $0.28^{+0.02}_{-0.02}$ & $0.30^{+0.14}_{-0.14}$\\ \hline
ngc6838 & $11.21^{+1.59}_{-1.59}$ & $-0.91^{+0.15}_{-0.13}$ & $4.16^{+0.21}_{-0.14}$ & $0.72^{+0.05}_{-0.04}$ & $0.16^{+0.14}_{-0.18}$\\ \hline
ngc6934 & $13.24^{+0.71}_{-0.71}$ & $-1.54^{+0.08}_{-0.14}$ & $15.99^{+0.39}_{-0.29}$ & $0.28^{+0.02}_{-0.03}$ & $0.21^{+0.12}_{-0.18}$\\ \hline
ngc6981 & $12.72^{+0.69}_{-0.69}$ & $-1.47^{+0.12}_{-0.14}$ & $17.08^{+0.49}_{-0.32}$ & $0.12^{+0.02}_{-0.02}$ & $0.22^{+0.14}_{-0.20}$\\ \hline
ngc7006 & $13.18^{+1.14}_{-1.00}$ & $-1.51^{+0.12}_{-0.18}$ & $39.78^{+2.11}_{-1.41}$ & $0.24^{+0.03}_{-0.03}$ & $0.03^{+0.10}_{-0.20}$\\ \hline
ngc7078 & $13.28^{+0.82}_{-0.71}$ & $-2.36^{+0.08}_{-0.13}$ & $11.25^{+0.22}_{-0.33}$ & $0.25^{+0.02}_{-0.03}$ & $0.18^{+0.14}_{-0.20}$\\ \hline
ngc7089 & $13.08^{+0.85}_{-0.85}$ & $-1.65^{+0.14}_{-0.14}$ & $12.05^{+0.35}_{-0.29}$ & $0.14^{+0.03}_{-0.03}$ & $0.18^{+0.16}_{-0.16}$\\ \hline
ngc7099 & $12.82^{+0.33}_{-0.50}$ & $-2.22^{+0.12}_{-0.14}$ & $8.96^{+0.16}_{-0.13}$ & $0.14^{+0.02}_{-0.03}$ & $0.34^{+0.14}_{-0.20}$\\ \hline
palomar1 & $8.20^{+3.87}_{-1.93}$ & $-0.69^{+0.17}_{-0.17}$ & $11.44^{+0.78}_{-0.78}$ & $0.47^{+0.05}_{-0.05}$ & $-0.05^{+0.16}_{-0.12}$\\ \hline
palomar12 & $9.94^{+0.92}_{-0.73}$ & $-0.86^{+0.13}_{-0.13}$ & $18.62^{+0.63}_{-0.42}$ & $0.08^{+0.03}_{-0.04}$ & $-0.01^{+0.14}_{-0.14}$\\ \hline
palomar15 & $13.97^{+0.88}_{-1.76}$ & $-2.06^{+0.19}_{-0.19}$ & $49.35^{+4.04}_{-4.04}$ & $1.21^{+0.03}_{-0.04}$ & $-0.01^{+0.14}_{-0.16}$\\ \hline
pyxis & $14.84^{+0.00}_{-3.28}$ & $-1.13^{+0.22}_{-0.15}$ & $38.29^{+2.77}_{-3.46}$ & $0.70^{+0.05}_{-0.04}$ & $0.12^{+0.14}_{-0.20}$\\ \hline
ruprecht106 & $11.30^{+1.55}_{-1.55}$ & $-1.62^{+0.13}_{-0.13}$ & $21.95^{+0.73}_{-0.73}$ & $0.56^{+0.03}_{-0.03}$ & $0.07^{+0.14}_{-0.16}$\\ \hline
terzan7 & $8.10^{+1.96}_{-1.40}$ & $-0.53^{+0.05}_{-0.14}$ & $23.65^{+1.42}_{-1.13}$ & $0.21^{+0.05}_{-0.05}$ & $-0.08^{+0.09}_{-0.11}$\\ \hline
terzan8 & $13.48^{+0.90}_{-0.77}$ & $-2.22^{+0.16}_{-0.12}$ & $28.74^{+0.99}_{-0.79}$ & $0.35^{+0.03}_{-0.03}$ & $0.34^{+0.18}_{-0.16}$\\ \hline
\end{tabular}%
\caption{}
\label{tab:my-table2}
\end{table}

\clearpage
\section{Fits to ACS globular clusters}
\label{app:fits}
In this appendix we show fits for typical Globular clusters in the ACS sample as an illustration of the adopted methodology. For each GC the upper row of Figure~\ref{fig:GCC1}  shows the color-magnitude diagram for the globular cluster. The gray points correspond to the individual stars, the red points show the best fit isochrone for the \texttt{DSED} model.  The bottom row shows the marginalized posteriors of the model parameters obtained applying our analysis.  The contours indicate the two-dimensional  68\%, 95\% and 99.7\% confidence levels constraints, while the panels in the diagonal show the one-dimensional marginalized posteriors. \\\

We also show the comparison between the best fit of the 68 Globular clusters and spectroscopic values from Dotter et al. catalog \citep{Dotter2010}. We find a very good agreement. For alsmost all the clusters, the spectroscopic value is within the 1-$\sigma$ range of the best fit.

\begin{figure}[h!]
 \begin{centering}
\includegraphics[scale=0.59]{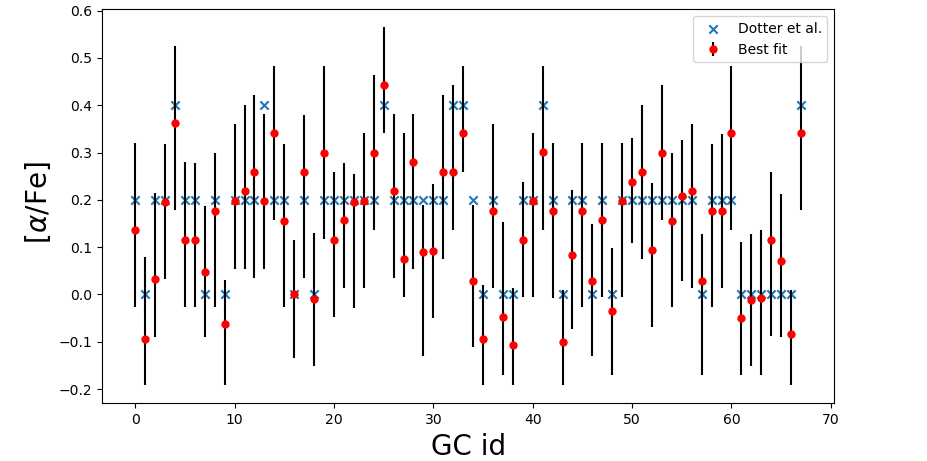}
\caption{Comparison of [$\alpha$/Fe] best fits and 1-$\sigma$ errors with spectroscopic measurements from Dotter et al \citep{Dotter2010}.}  
\label{fig:GCC1}
\end{centering}
\end{figure}

\begin{figure}[h!]
 \begin{centering}
\includegraphics[width=0.33\textwidth]{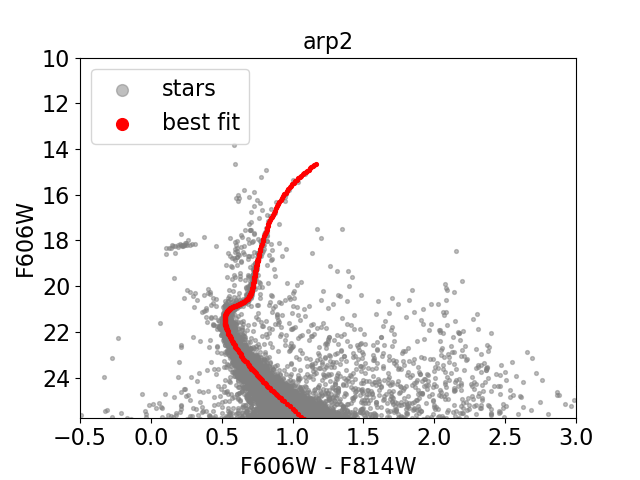}\linebreak[0]%
\includegraphics[width=0.33\textwidth]{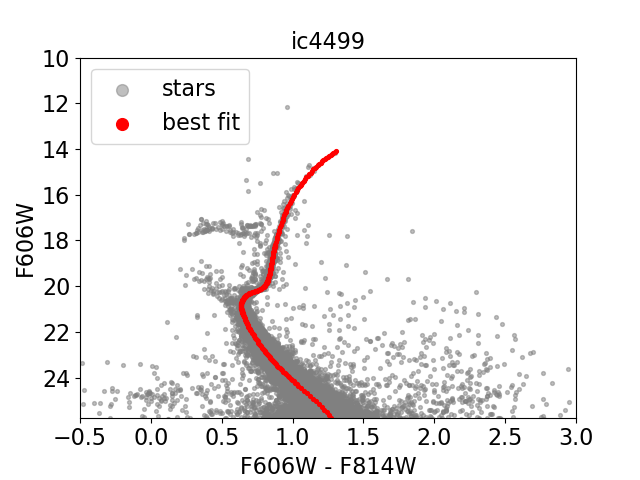}\linebreak[0]%
\includegraphics[width=0.33\textwidth]{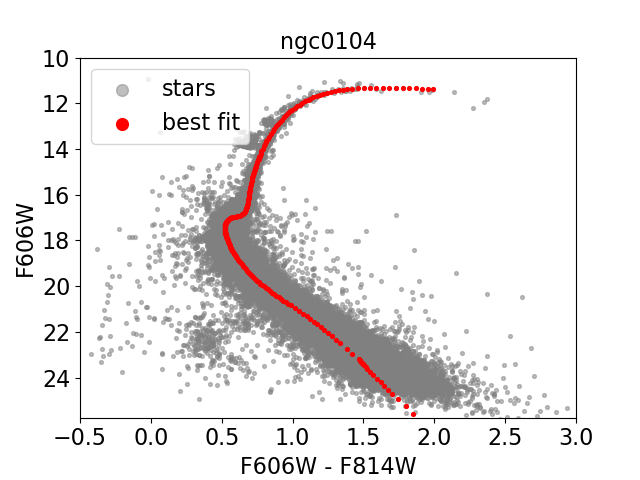}\linebreak[0]%
\includegraphics[width=0.33\textwidth]{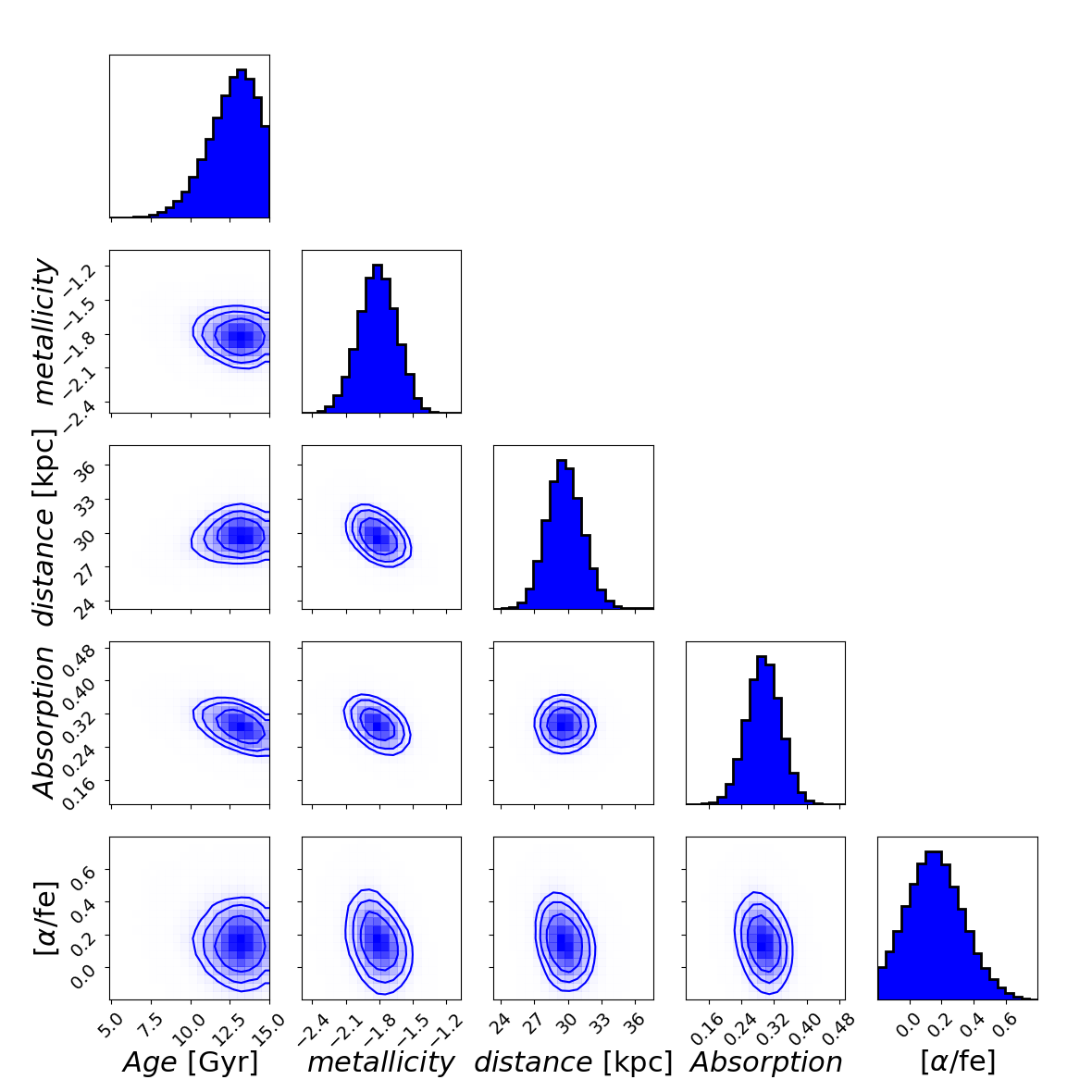}\linebreak[0]%
\includegraphics[width=0.33\textwidth]{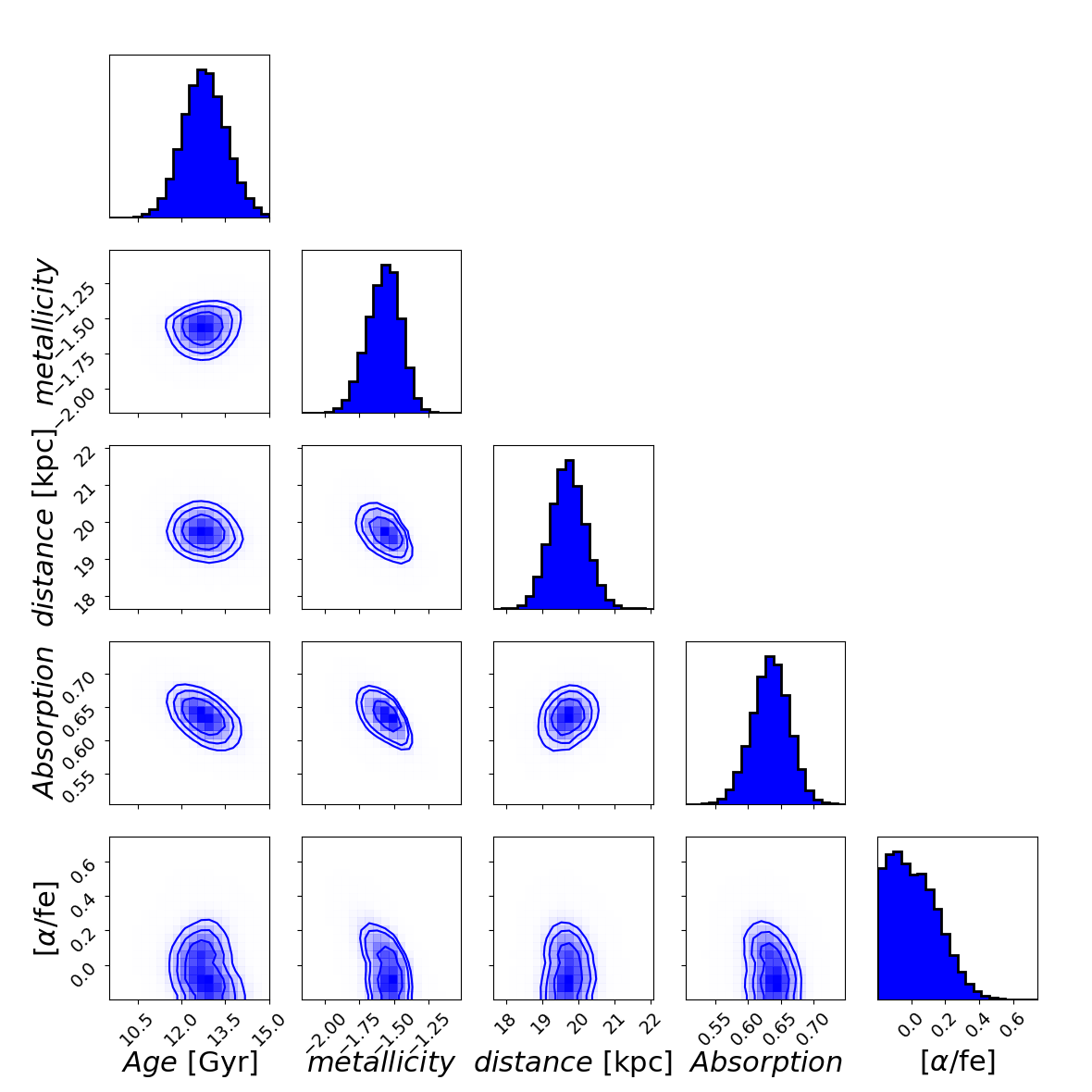}\linebreak[0]%
\includegraphics[width=0.33\textwidth]{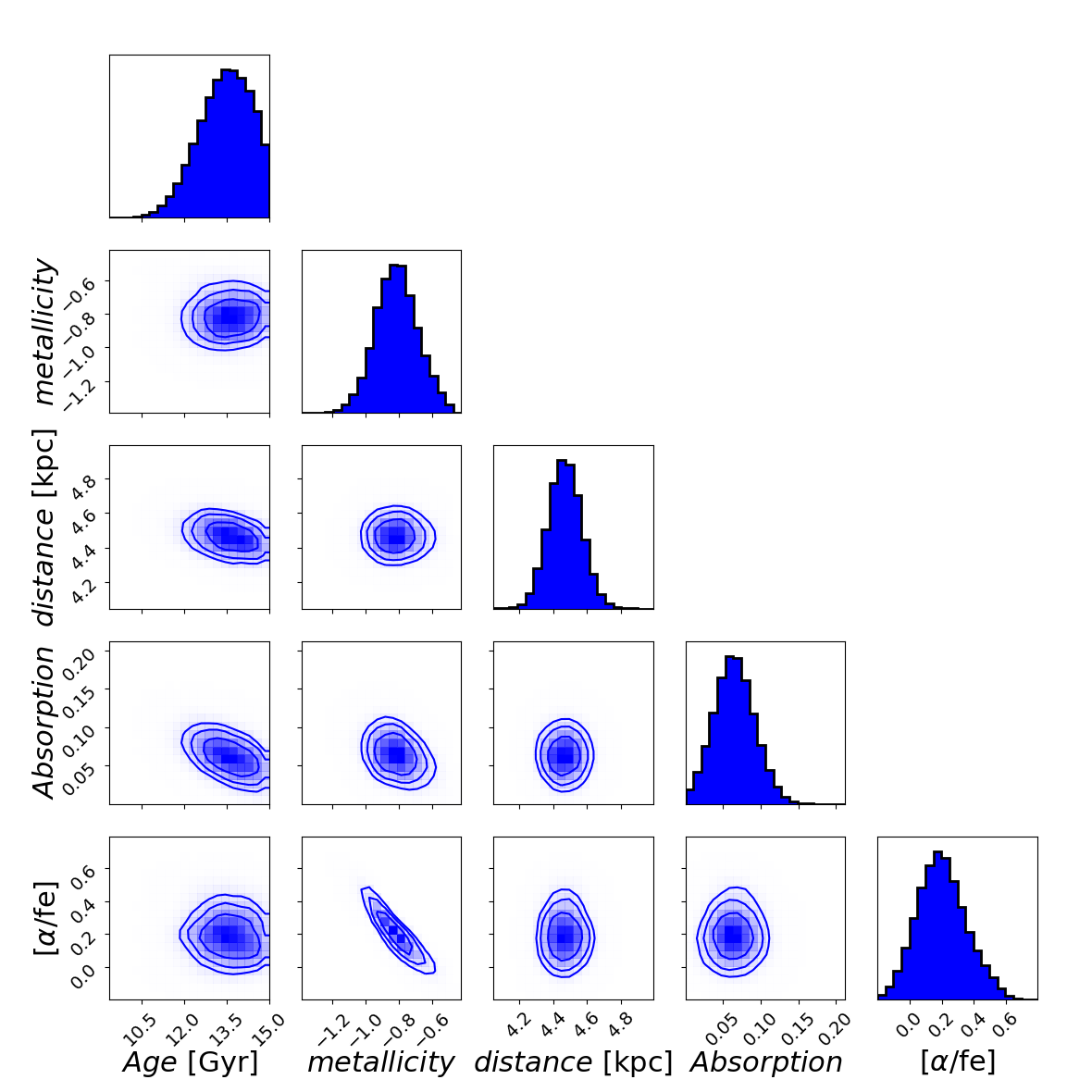}\linebreak[0]%

\includegraphics[width=0.33\textwidth]{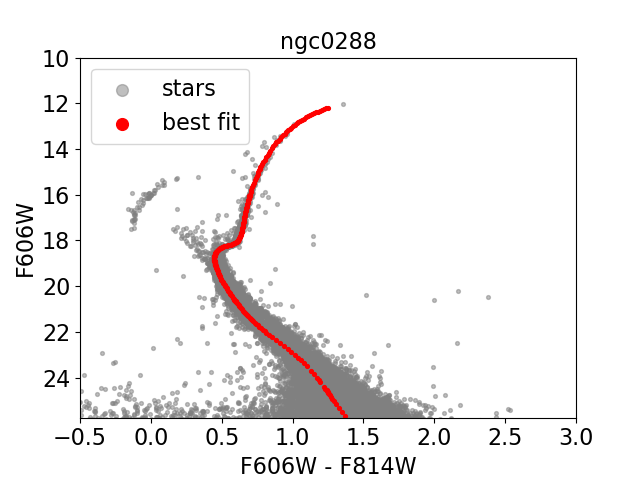}\linebreak[0]%
\includegraphics[width=0.33\textwidth]{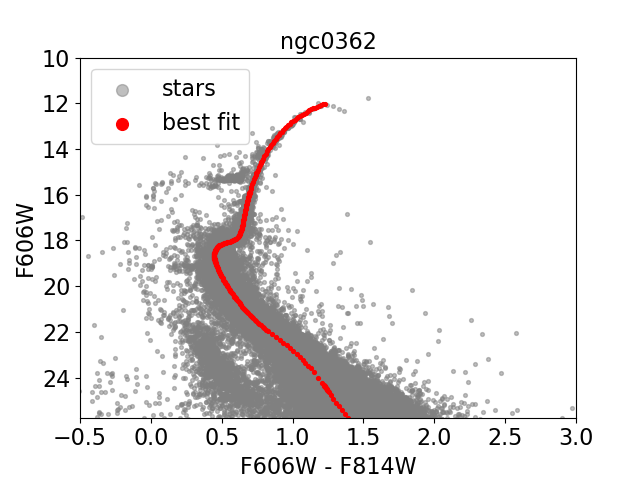}\linebreak[0]%
\includegraphics[width=0.33\textwidth]{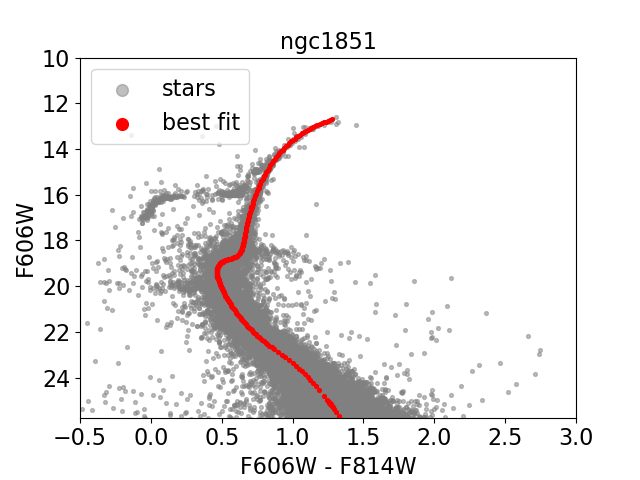}\linebreak[0]%
\includegraphics[width=0.33\textwidth]{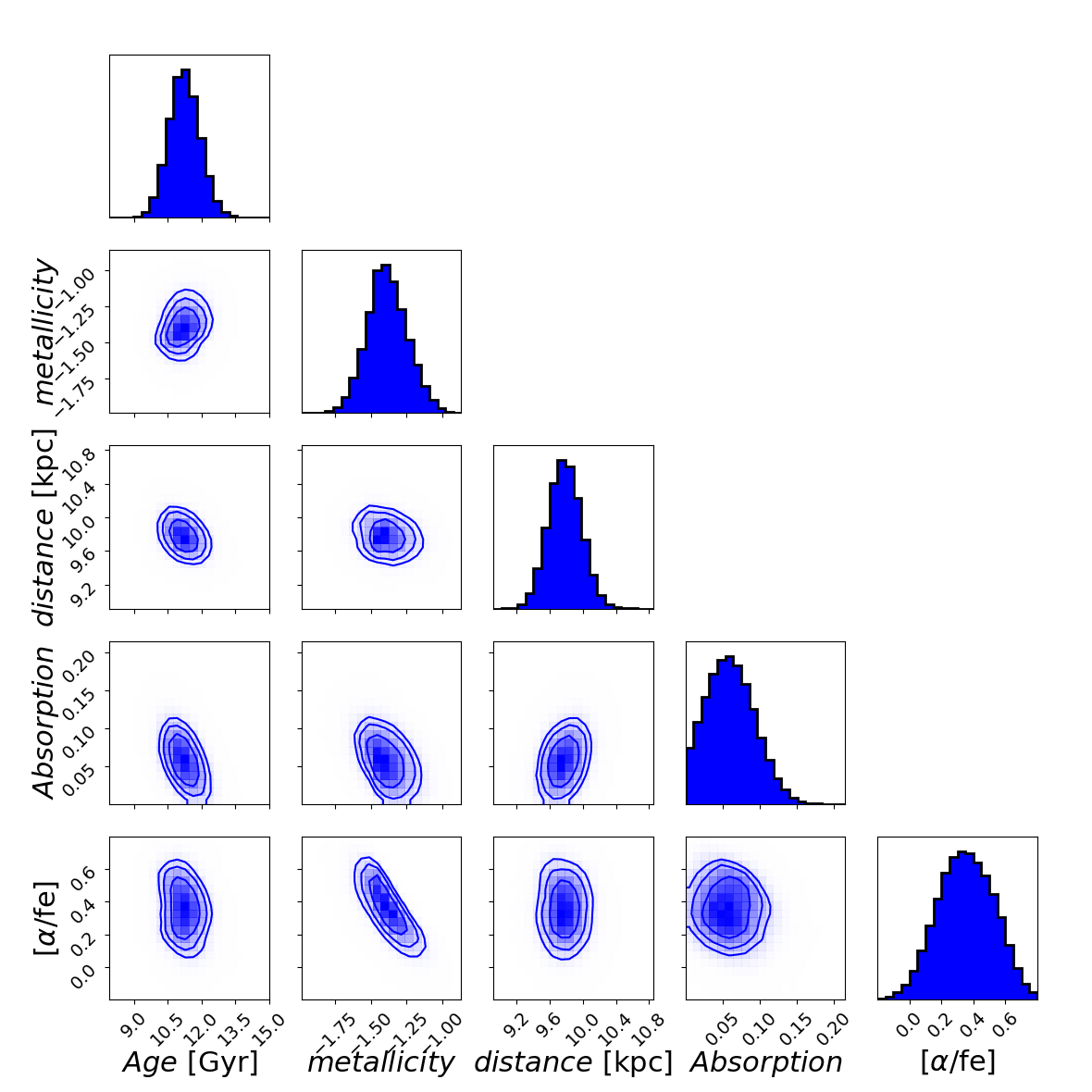}\linebreak[0]%
\includegraphics[width=0.33\textwidth]{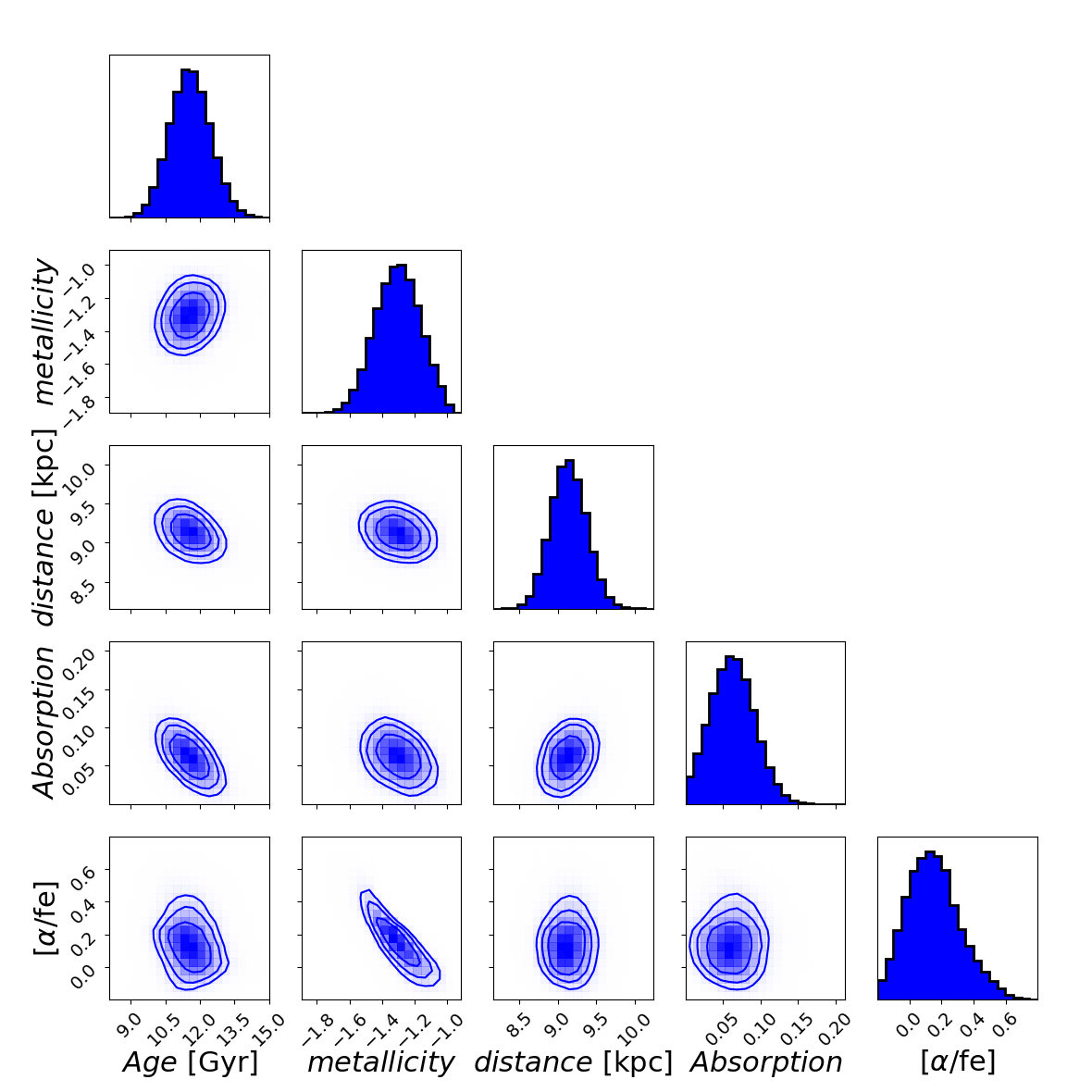}\linebreak[0]%
\includegraphics[width=0.33\textwidth]{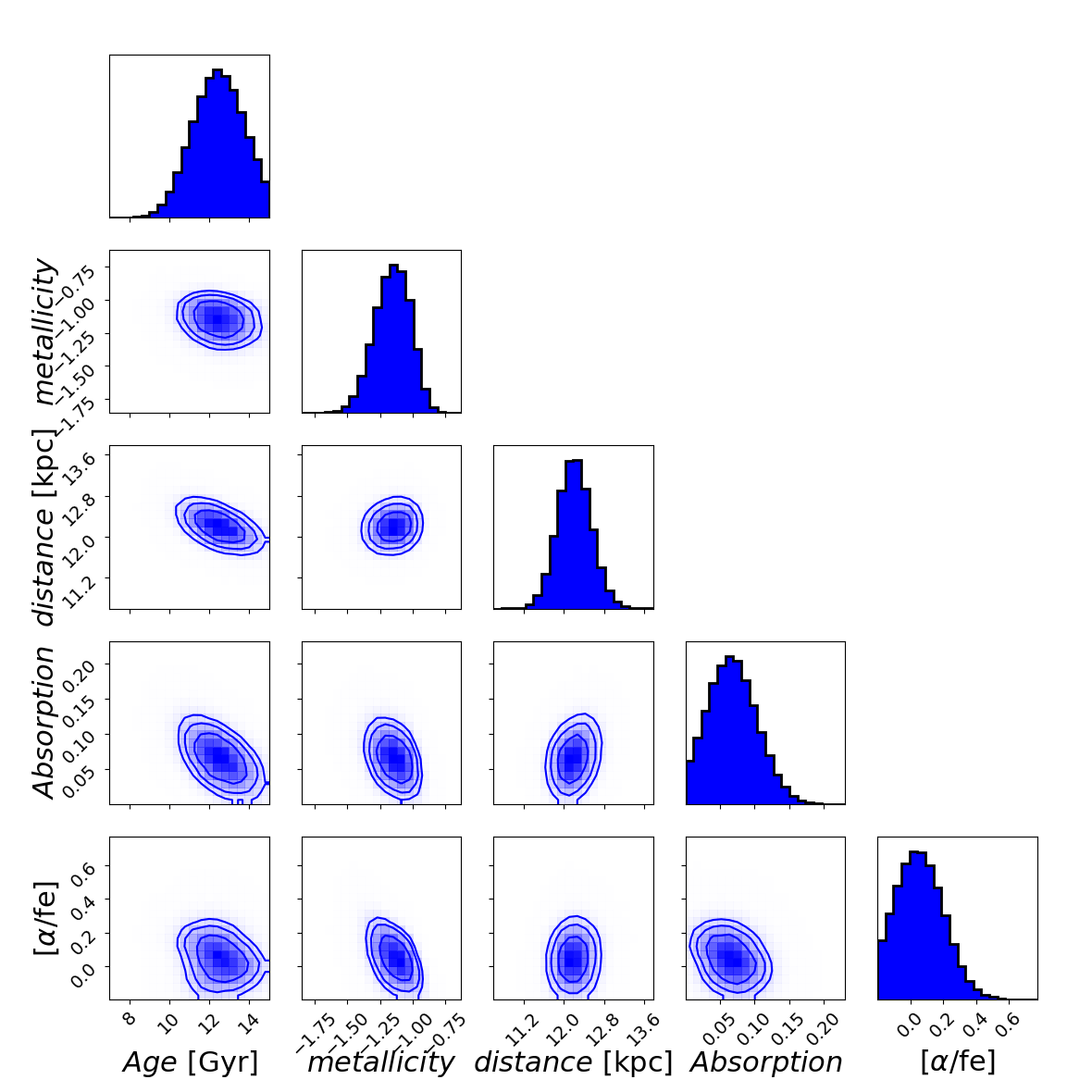}\linebreak[0]%

\caption{Fit to 6 Globular clusters and the corresponding join Bayesian posterior for the corresponding parameters. The contour levels are confidence, 2D join, intervals for $1, 2$ and $3-\sigma$.}  
\label{fig:GCC1}
\end{centering}
\end{figure}

\clearpage
\section{Fitting formula for the distribution of $\Delta_t$}
\label{appendix:fitDt}
\begin{figure}[h!]
\centering
\includegraphics[scale=1]{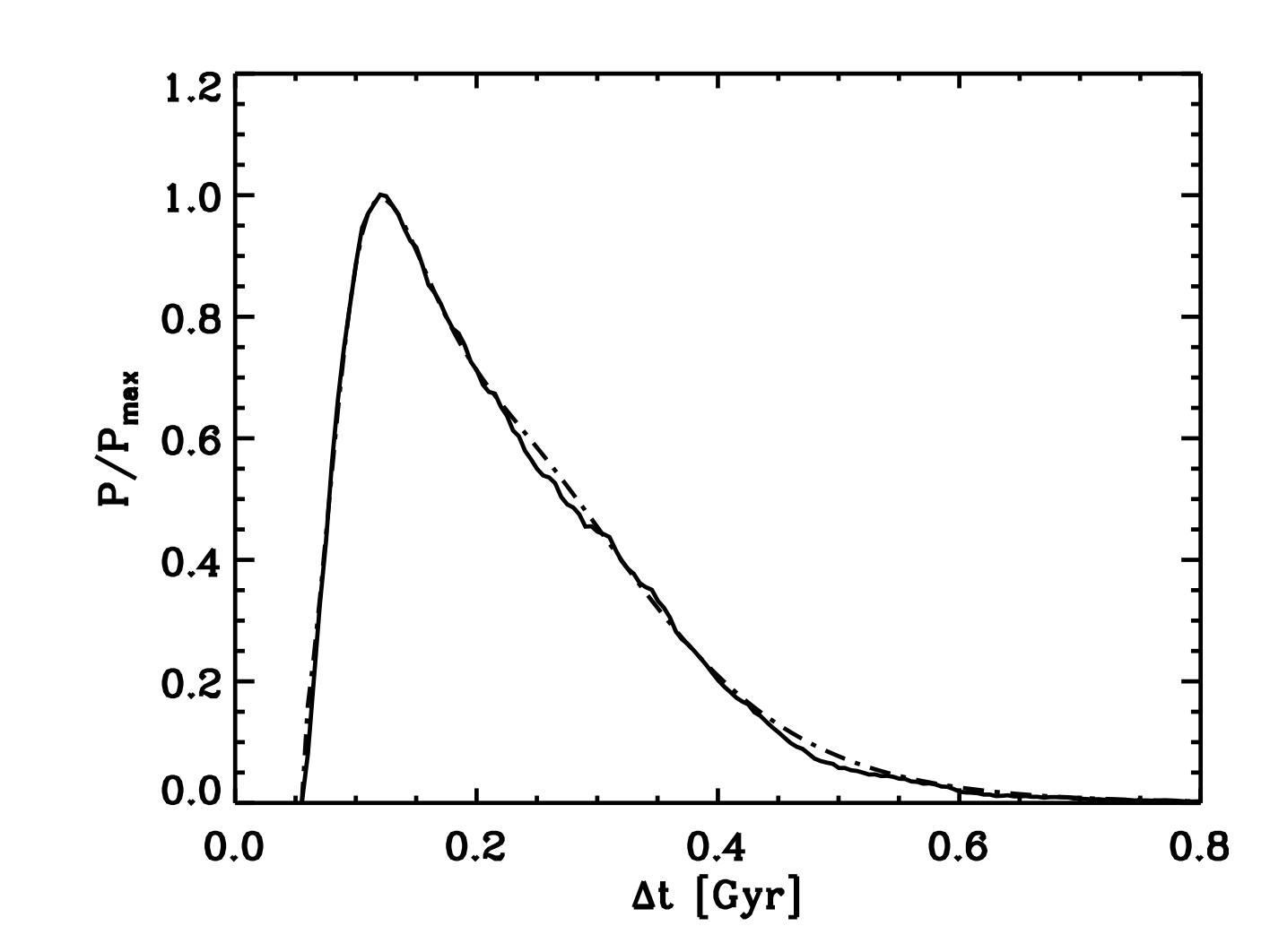} 
\caption{Distribution of the  $\Delta_t$ taken from the right panel of Figure 1 of Ref.~\cite{JimGC} (solid line) and fitting formula used here (dot-dashed line).}
\label{Dtdistr}
\end{figure}

The distribution of $\Delta_t$ shown in  the right panel of Figure 1 of Ref.~\cite{JimGC} can be well approximated by the following fitting formula (see Fig~\ref{Dtdistr}). Let $x$ indicate $\Delta_t$,  $l=\log_{10} (\Delta_t)$ and $l_1 \equiv \log_{10}(0.1155)$, $l_2 \equiv \log_{10}(0.255)$, $\sigma_1 =0.15$, $\sigma_1'=0.17$, $\sigma_2=0.155$ then 
\begin{eqnarray}
F_1(x)&=&\exp\left(-\frac{1}{2}\frac{(l-l_1)^2}{\sigma_1^2}\right) \,\,\,\ {\rm if} \,x \le 0.1155  \\
F_1(x)&=&\exp\left(-\frac{1}{2}\frac{(l-l_1)^2}{\sigma_1'^2}\right) \,\,\,\ {\rm if} \,x \ge 0.1155  \\
F_2(x)&=&\exp\left(-\frac{1}{2}\frac{(l-l_2)^2}{\sigma_2^2}\right) \\
P_{\Delta_t}(x)&\propto & 0.95\,F_1(x)+0.45\, F_2(x)
\end{eqnarray}

\end{document}